\documentclass[onecolumn]{emulateapj}
\usepackage{epsf}
\slugcomment{}
\shortauthors{Taruya \& Hiramatsu}
\shorttitle{Closure theory for cosmological power spectra}
\newcommand{\simgt}{\lower.5ex\hbox{$\; \buildrel > \over \sim \;$}}
\newcommand{\simlt}{\lower.5ex\hbox{$\; \buildrel < \over \sim \;$}}
\newcommand{\bfv}{\mbox{\boldmath$v$}}
\newcommand{\bfx}{\mbox{\boldmath$x$}}
\newcommand{\bfk}{\mbox{\boldmath$k$}}
\newcommand{\bfp}{\mbox{\boldmath$p$}}
\newcommand{\bfq}{\mbox{\boldmath$q$}}

\newcommand{\tdPhi}{\widetilde{\Phi}}
\newcommand{\tdG}{\widetilde{G}}
\newcommand{\tdP}{\widetilde{P}}
\newcommand{\tdR}{\widetilde{R}}
\begin{document}

\title{A Closure Theory for Non-linear Evolution of 
Cosmological Power Spectra}

\author{Atsushi Taruya, Takashi Hiramatsu} 

\affil{Research Center for the Early Universe (RESCEU), 
School of Science, The University of Tokyo, Tokyo 113-0033, Japan.}

\begin{abstract}
We apply a non-linear statistical method in turbulence to 
the cosmological perturbation theory and derive a closed set of 
evolution equations for matter power spectra. The resultant closure 
equations consistently recover the one-loop results of standard 
perturbation theory and beyond that, 
it is still capable of treating the non-linear 
evolution of matter power spectra. We find 
the exact integral expressions for the solutions of closure equations. 
These analytic expressions coincide with 
the renormalized one-loop results presented 
by \citet{CS2006a,CS2007}, apart from the vertex renormalization. 
By constructing 
the non-linear propagator, we analytically evaluate the non-linear matter 
power spectra based on the first-order Born approximation of the 
integral expressions and compare it with those of the renormalized 
perturbation theory. 
\end{abstract}

\section{Introduction}
\label{sec:introduction}

Cosmology now enters the era of precision cosmology. With large data 
set from the precision measurements of 
the cosmic microwave background anisotropies as well as 
the matter density fluctuations in the large-scale structure, 
the standard cosmological model has been fully established 
\citep[e.g.,][]{S2006,T2006}.  
The associated cosmological parameters are well-determined with 
an accuracy of less than $10\%$ level. 
With improved sensitivity and higher precision of future observations,  
the modern picture of the universe will be further reinforced  
and one can even explore 
a tiny signature of new physics beyond the standard cosmological model.

In fact, several ambitious missions for galaxy redshift survey 
are planed in order to reveal the nature of dark energy 
\citep[e.g.,][and references therein]{A2006,P2006}. Among these, 
Wide-field Fiber-fed Multi-Object Spectrograph (WFMOS) may be 
one of the best facility capable of achieving the percent-level 
measurement of baryon acoustic oscillations (BAOs) imprinted in the 
matter power spectrum \citep[][]{MWP1999}. The recent observations from 
Sloan Digital Sky Survey (SDSS) and 
Two-degree-Field Galaxy Redshift Survey (2dF GRS) showed 
that the characteristic 
scale of BAOs can be used as the cosmic standard ruler to determine 
the distance-redshift relation of high-redshift galaxies 
\citep[][]{C2005,E2005,H2006,P2007}. 
Since the distance-redshift relation is sensitive to the cosmic 
expansion history, details of the accelerated expansion can be 
clarified from the accurate measurement of BAOs \citep[][]{SE2003}. 
With percent-level measurement of characteristic 
scale of BAOs, the determination of dark energy equation-of-state, 
parametrized by $w\equiv P_{\rm de}/\rho_{\rm de}$, 
will achieve a few percent accuracy, where 
$P_{\rm de}$ and $\rho_{\rm de}$ are the pressure and the energy 
density of dark energy, respectively.

On the other hand, pursuit of the nature of dark energy highlights 
various fundamental problems which are potentially crucial 
for the accurate determination of dark energy equation-of-state. 
For example, the observation of BAOs requires high-precision theoretical 
template for the matter power spectrum in the relevant wavenumber, 
$k\sim0.1-0.3h$Mpc$^{-1}$. To achieve the required accuracy 
in the determination of $w$, several systematic effects 
must be incorporated into the theoretical predictions.  
Among known systematic effects, the non-linear gravitational clustering 
is one of the most fundamental building block in the theory of 
structure formation. The recent $N$-body simulations showed that 
the non-linear growth of matter distribution significantly alters  
the shape of the power spectrum and the acoustic signature of BAOs 
tends to be erased, where linear theory prediction of matter power 
spectrum is no longer valid \citep[e.g.,][]{MWP1999,SE2005}. To tackle 
the issue, the perturbation theory for gravitational clustering has been 
revived and has been applied to the study of BAOs 
\citep[e.g.,][]{SS1991,MSS1992,JB1994,SF1996,JK2006,N2007}. 
The inclusion of leading-order correction to the 
non-linear clustering effect somehow improves a performance and 
reproduces the $N$-body results very well \citep[][]{JK2006}. 
At lower redshifts $z<2$, 
however, next-to-leading order effect becomes important and the 
theoretical prediction with leading-order correction 
is insufficient to reproduce the $N$-body simulations.

Going beyond the perturbation theory, 
existing theoretical tools dealing with the non-linear 
gravitational clustering are the $N$-body simulation and the fitting 
formula for matter power spectrum \citep[e.g.,][]{PD1996,S2003}, 
as well as the phenomenological approach 
based on the halo model \cite[][for a review]{CS2002}. 
Currently, however, none of the reliable methods 
to ensure the percent-level precision exist. 
While the $N$-body simulation has a potential to provide a 
high-precision prediction, at present, one cannot blindly trust the 
$N$-body results unless reliable and comparable counterpart 
is established and is fully reconciled with $N$-body results. 
In this respect, development of new analytical method beyond 
the perturbation theory is necessary and essential for the 
progress on precision cosmology.

In this paper, we present a non-linear statistical method to predict 
the time evolution of matter power spectrum. Very recently, there appear  
several works on the statistical treatment going beyond the perturbation 
theory \citep[][]{V2004,CS2006a,CS2006b,M2006,V2007a,MP2007,IS2007}. 
Based on the field-theoretical approach, 
the perturbation theory has been reformulated 
by improving the summation of the naive perturbative expansion.  The 
so-called {\it renormalized perturbation theory} (RPT) developed by 
\citet{CS2006a} seems viable 
theoretical tool alternative to the $N$-body simulation, suited to a 
high-precision prediction. Using RPT, some attempts to predict 
the non-linear evolution of BAOs has been reported \citep[][]{CS2007}. 
Here, we consider an alternative statistical method 
accepted widely in the subject of statistical theory of turbulence 
\citep[e.g.,][]{K1959,L1973,K1981,KG1997}. 
In contrast to the sophisticated treatment based on the 
field-theoretical approach, 
our method is rather primitive in the sense that the effect of non-linearity  
of matter power spectrum is simply described 
by a systematic expansion and a truncation of the naive perturbation.  
After applying the so-called reversed expansion, 
the perturbative expansion is effectively re-organized and 
a class of higher-order corrections is systematically 
re-summed \citep{W1961}. With 
this treatment, some non-perturbative effects are also incorporated. 
We derive a closed set of moment equations characterizing the 
non-linear evolution of power spectra. The resultant evolution equations 
consistently recover the leading-order results of standard 
perturbation theory (one-loop PT). The solutions for 
these closure equations have exact integral expressions, which  
coincides with the one-loop results of RPT.  
Constructing the non-linear propagator, we attempt to evaluate 
the non-linear matter power spectra analytically. 
Based on the first-order Born approximation of the exact integral 
expressions, we find that the power spectra from the 
closure theory reasonably agree with those from the RPT.

The outline of the paper is as follows. 
In Section \ref{sec:preliminaries}, 
governing equations for the dynamics of cosmological gravitational clustering 
is presented on the basis of the fluid description of the Vlasov equation. 
The {\it closure problem}, i.e., 
theoretical issue on the self-consistent treatment of the hierarchy of 
moment equations, is briefly mentioned. 
In Section \ref{sec:closure_theory}, 
non-linear statistical method in turbulence called 
{\it direct interaction approximation} is introduced and is applied to 
the present cosmological situation. Then, a closed set of evolution 
equations for matter power spectrum is obtained 
by the systematic perturbative expansion and the so-called reversion 
procedure. In Section \ref{sec:properties}, important properties of 
the resultant closure equations, i.e., recovery of standard PT and exact 
integral solutions, are discussed, based on detailed mathematical calculations 
presented in Appendices \ref{appendix:one-loop_PT} and 
\ref{appendix:integral_sol}. In Section \ref{sec:analytic_treatment},  
analytical treatment for our closure system is presented for illustrative 
purpose. Constructing approximate solutions for non-linear propagator, 
power spectrum of density fluctuations is calculated 
based on the Born approximation of the integral solutions. 
The results are compared with those obtained from RPT, particularly 
focusing on the non-linear evolution of BAOs. 
Finally, Section \ref{sec:conclusion} is devoted to discussion and 
conclusions.

\section{Preliminaries}
\label{sec:preliminaries}

\subsection{Basic equations}
\label{subsec:basic_eqs}

Throughout the paper, we consider the evolution of mass distribution in the 
flat universe, neglecting the tiny contribution from the massive neutrinos. 
To evaluate the non-linear growth of the density perturbations, 
hydrodynamic description of mass distribution is useful. Strictly speaking, 
this treatment is not exact and is often called the single-stream 
approximation of the Vlasov equation. However, at least in a statistical 
sense, it would be the best approximation if the scale of our interest is 
sufficiently large so that one can safely ignore the effect of 
shell-crossing.  Denoting the fluctuation of mass distribution 
consisting of the cold dark matter and the baryon fluid by $\delta$, we 
have
\begin{eqnarray}
&&\frac{\partial\delta}{\partial t}+\frac{1}{a}\,\nabla\{(1+\delta)\bfv\}=0,
\label{eq:eq_continuity0}
\\
&&\frac{\partial\bfv}{\partial t}+H\bfv+
\frac{1}{a}\,(\bfv\cdot\nabla)\bfv=-\frac{1}{a}\,\nabla\phi,
\label{eq:Euler_eq}
\\
&&\frac{1}{a^2}\,\nabla^2\phi=4\pi\,G\,\rho_{\rm m}\,\delta
\label{eq:Poisson_eq}
\end{eqnarray}
Here, $a$ is the scale factor of the universe, $H$ is the Hubble parameter 
and $\rho_{\rm m}$ is the homogeneous mass density field. 
Assuming the irrotationality of the fluid flow, 
the above equations can be recast as \citep[e.g.,][]{Taruya2000}: 
\begin{eqnarray}
&&\frac{\partial\delta}{\partial t}+
H\,\theta+\frac{1}{a}\nabla(\delta\cdot \bfv)=0,
\label{eq:eq_continuity}
\\
&&\nonumber
\\
&&\frac{\partial \theta}{\partial t}+
\frac{1}{2}\left(1-3w\,\Omega_w\right)H\,\theta + 
\frac{1}{a^2H}\,\nabla\cdot\left\{(\bfv\cdot\nabla)\bfv\right\}+
\frac{3}{2}H(1-\Omega_w)\delta = 0, 
\label{eq:euler_eq}
\end{eqnarray}
where the quantity $\theta$ is the velocity divergence defined by 
\begin{equation}
\theta = \frac{\nabla\cdot \bfv}{aH}. 
\end{equation}
The quantity $\Omega_w$ is the density parameter of dark energy 
satisfying the equation of sate, $P_{\rm de}=w\rho_{\rm de}$, 
defined by $\Omega_w=8\pi\,G\,\rho_{\rm de}/(3H^2)$. Note that 
the relation $\Omega_{\rm m} +\Omega_{w}=1$ holds in the flat 
cosmology.

To treat the non-linear evolution of the matter power spectrum, 
the Fourier representation of equations (\ref{eq:eq_continuity}) 
and (\ref{eq:euler_eq}) is useful. To do this, we introduce 
the Fourier transform of the perturbed quantities: 
\begin{equation}
\delta(\bfx;t)
=\int\frac{d^3\bfk}{(2\pi)^3}\,e^{-i\,\bfk\cdot \bfx}\,
\delta(\bfk;t),
\quad
\theta(\bfx;t)=\int\frac{d^3\bfk}{(2\pi)^3}\,e^{-i\,\bfk\cdot \bfx}\,
\theta(\bfk;t).
\end{equation}
The assumption of irrotational flow implies
\begin{equation}
\bfv(\bfx;t)=\int\frac{d^3\bfk}{(2\pi)^3}\,e^{-i\,\bfk\cdot \bfx}\,
\frac{i\,\bfk}{|\bfk|^2}\,\widetilde{\theta}(\bfk;t).
\end{equation}
Then, the fluid equations (\ref{eq:eq_continuity}) and (\ref{eq:euler_eq}) 
can be written as \citep[e.g.,][]{Taruya2000}:
\begin{eqnarray}
&&H^{-1}\frac{\partial\delta(\bfk;t)}{\partial t} + 
\theta(\bfk;t) =\,-\int\frac{d^3\bfk'}{(2\pi)^3}\,
\alpha(\bfk',\bfk-\bfk')\,\delta(\bfk-\bfk';t)\,
\theta(\bfk';t),
\label{eq:perturb1}
\\
&&\nonumber
\\
&&H^{-1}\frac{\partial\theta(\bfk;t)}{\partial t} + 
\frac{1}{2}(1-3w\Omega_w)\theta(\bfk;t) 
+\frac{3}{2}(1-\Omega_w)\delta(\bfk;t) 
\nonumber
\\
&&\quad\quad\quad\quad\quad\quad\quad\quad\quad~~~
=\,-\frac{1}{2}\,\int\frac{d^3\bfk'}{(2\pi)^3}\,
\beta(\bfk',\bfk-\bfk')\,\theta(\bfk-\bfk';t)\,
\theta(\bfk';t),
\label{eq:perturb2}
\end{eqnarray}
where the kernels in the Fourier integrals, $\mathcal{\alpha}$  and 
$\mathcal{\beta}$,  are respectively given by 
\begin{eqnarray}
\alpha(\bfk_1,\bfk_2)=1+\frac{\bfk_1\cdot\bfk_2}{|\bfk_1|^2},
\quad\quad
\beta(\bfk_1,\bfk_2)=
\frac{(\bfk_1\cdot\bfk_2)\left|\bfk_1+\bfk_2\right|^2}{|\bfk_1|^2|\bfk_2|^2}.
\end{eqnarray}

For later analysis, it is convenient to introduce 
the vector-field notation 
and rewrite the equations (\ref{eq:perturb1}) and (\ref{eq:perturb2}) in 
more compact form. For this purpose, we define
\begin{equation}
\Phi_a(\bfk;t)=\left(
\begin{array}{c}
\delta(\bfk;t) \\
-\theta(\bfk;t)/f(t)
\end{array}
\right),  
\end{equation}
where the function $f(t)$ is given by $f(t)\equiv d\ln D(t)/d\ln a$ and 
the quantity $D(t)$ being the linear growth rate.  
Then, in terms of the new time variable $\eta\equiv\ln D(t)$, the 
evolution equation for the vector quantity $\Phi_a(\bfk;t)$ becomes 
\citep[e.g.,][]{CS2006a,V2007a}: 
\begin{equation}
\left[\delta_{ab}\,\frac{\partial}{\partial \eta}+\Omega_{ab}(\eta)\right]
\Phi_b(\bfk;t)=\int\frac{d^3\bfk_1\,d^3\bfk_2}{(2\pi)^3}
\delta_D(\bfk-\bfk_1-\bfk_2)\,\gamma_{abc}(\bfk_1,\bfk_2)\,
\Phi_b(\bfk_1;t)\,\Phi_c(\bfk_2;t),
\label{eq:vec_fluid_eq}
\end{equation}
where $\delta_D$ is the Dirac delta function. 
Here and in what follows, 
we use the summation convention that the repetition of the same 
subscripts  indicates the sum over the whole vector components. 
The time-dependent matrix $\Omega_{ab}(\eta)$ is given by
\begin{equation}
\Omega_{ab}(\eta)=\left(
\begin{array}{cc}
{\displaystyle 0} & {\displaystyle -1 }
\\
{\displaystyle -\frac{3}{2f^2}(1-\Omega_w)} \quad&\quad 
{\displaystyle \frac{3}{2f^2}\Omega_{\rm m}-1}
\end{array}
\right).
\label{eq:matrix_M}
\end{equation}
Each component of the vertex function $\gamma_{abc}$ becomes
\begin{eqnarray}
\gamma_{abc}(\bfk_1,\bfk_2) =
\left\{
\begin{array}{ccl} 
\alpha(\bfk_2,\bfk_1)/2 &;& (a,b,c)=(1,1,2) 
\\
\alpha(\bfk_1,\bfk_2)/2 &;& (a,b,c)=(1,2,1) 
\\
\beta(\bfk_1,\bfk_2)/2 &;& (a,b,c)=(2,2,2) 
\\
     0                       &;&  \mbox{otherwise}
\end{array}
\right..
\label{eq:def_Gamma}
\end{eqnarray}
Note that the vertex function $\gamma_{abc}$ has the following 
symmetric properties: 
$\gamma_{abc}(\bfk_1,\bfk_2)=\gamma_{acb}(\bfk_2,\bfk_1)$, 
$\gamma_{abc}(-\bfk_1,-\bfk_2)=\gamma_{abc}(\bfk_1,\bfk_2)$, 
$\gamma_{abc}(\bfk_1,-\bfk_2)=\gamma_{abc}(-\bfk_1,\bfk_2)$ and 
 $\gamma_{abc}(\bfk,-\bfk)=0$. 
Equation (\ref{eq:vec_fluid_eq}) with (\ref{eq:matrix_M}) 
(\ref{eq:def_Gamma}) is the basic equation for our subsequent analysis.

\subsection{Moment equations}
\label{subsec:moment_eqs}

Before addressing the non-linear statistical method, 
it is worthwhile to mention the closure problem for the dynamics of 
statistical quantities.

First of all, we define the two kinds of the power spectra for 
fluid field $u_i(\bfk;t)$: 
\begin{eqnarray}
\langle \Phi_a(\bfk;\eta)\, \Phi_b(\bfk';\eta)\rangle 
&=& (2\pi)^3\,\delta_D(\bfk+\bfk')\,P_{ab}(|\bfk|;\,\eta),
\nonumber\\
\langle \Phi_a(\bfk;\eta)\, \Phi_b(\bfk';\eta')\rangle 
&=& (2\pi)^3\,\delta_D(\bfk+\bfk')\,R_{ab}(|\bfk|;\,\eta,\eta'),
\quad\quad(\eta>\eta'),  
\label{eq:def_of_Pij_Rij}
\end{eqnarray}
where the bracket $\langle\cdot\rangle$ stands for the ensemble average. 
The quantity $P_{ab}$ is the ordinary power spectra which 
we are interested in and we have the  
symmetry, $P_{ab}=P_{ba}$.  On the other hand, the quantity $R_{ab}$
represents the cross power spectrum between different times, 
which will be later important when we derive a closed set of equations. 
Note that $R_{ab}\neq R_{ba}$, in general.

Since we are specifically concerned with the time evolution of 
statistical quantities $P_{ab}(k;\eta)$ and $R_{ab}(k;\eta,\eta')$, 
rather than  focusing on equation (\ref{eq:vec_fluid_eq}), 
it seems convenient to treat the moment equations for these quantities. 
To derive the moment equation for $P_{ab}$, we first note that 
\begin{eqnarray}
\frac{\partial}{\partial\eta}\,
\Bigl\langle \Phi_a(\bfk;\eta)\,\Phi_b(\bfk';\eta)\Bigr\rangle &=&
\left\langle \frac{\partial \Phi_a(\bfk;\eta)}{\partial \eta}\,
\Phi_b(\bfk';\eta)\right\rangle + \left\langle 
\Phi_a(\bfk;\eta)\,\frac{\partial \Phi_b(\bfk';\eta)}{\partial \eta}
\right\rangle . 
\end{eqnarray}
With a help of equation (\ref{eq:vec_fluid_eq}), 
we eliminate the time derivative $\partial \Phi_a/\partial\eta$.   
Then, we obtain 
\begin{eqnarray}
&&\widehat{\mbox{\boldmath$\Sigma$}}_{abcd}(\eta)\,
\Bigl\langle \Phi_c(\bfk;t)\, \Phi_d(\bfk';t)\Bigr\rangle 
\quad\quad\quad\quad\quad
\nonumber
\\
&&\quad
=\int\frac{d^3\bfk_1\,d^3\bfk_2}{(2\pi)^3}\,
\Bigl[\,\delta_D(\bfk'-\bfk_1-\bfk_2)\,\gamma_{bpq}(\bfk_1,\bfk_2) 
\Bigl\langle \Phi_a(\bfk;\eta)\, \Phi_p(\bfk_1;\eta)\, 
\Phi_q(\bfk_2;\eta)\Bigr\rangle 
\Bigr.
\nonumber
\\
&&\quad\quad\quad\quad\quad\quad\quad~ +
\Bigl.\delta_D(\bfk-\bfk_1-\bfk_2)\,\gamma_{apq}(\bfk_1,\bfk_2) 
\Bigl\langle \Phi_b(\bfk';\eta)\, \Phi_p(\bfk_1;\eta)\, 
\Phi_q(\bfk_2;\eta)\Bigr\rangle
\Bigr].
\label{eq:moment_Pij}
\end{eqnarray}
Similarly, moment equations for $R_{ab}$ becomes
\begin{eqnarray}
&& \widehat{\mbox{\boldmath$\Lambda$}}_{ab}(\eta)\,
\Bigl\langle \Phi_b(\bfk;\eta)\, \Phi_c(\bfk';\eta')\Bigr\rangle 
\quad\quad\quad
\nonumber
\\
&&\quad\quad
=\int\frac{d^3\bfk_1\,d^3\bfk_2}{(2\pi)^3}\,
\,\delta_D(\bfk-\bfk_1-\bfk_2)\,\gamma_{apq}(\bfk_1,\bfk_2) 
\Bigl\langle \Phi_c(\bfk';\eta')\, \Phi_p(\bfk_1;\eta)\, 
\Phi_q(\bfk_2;\eta)\Bigr\rangle.
\label{eq:moment_Rij}
\end{eqnarray}
Here, we have introduced the two kinds of operators, 
$\widehat{\mbox{\boldmath$\Sigma$}}_{abcd}$ and 
$\widehat{\mbox{\boldmath$\Lambda$}}_{ac}$: 
\begin{equation}
\widehat{\mbox{\boldmath$\Sigma$}}_{abcd}(\eta)\equiv
\delta_{ac}\delta_{bd}\,\frac{\partial}{\partial \eta}+
\delta_{ac}\,\Omega_{bd}(\eta) + \delta_{bd}\,\Omega_{ac}(\eta),
\quad\quad\quad
\widehat{\mbox{\boldmath$\Lambda$}}_{ab}(\eta)\equiv
\delta_{ab}\,\frac{\partial}{\partial \eta}+\Omega_{ab}(\eta) .
\label{eq:def_of_Sigma_Lambda}
\end{equation}

Equations (\ref{eq:moment_Pij}) and (\ref{eq:moment_Rij}) 
are not yet closed because they contain 
the higher-order correlation functions (or bi-spectra). 
In order to obtain the closed set of evolution equations, 
it is necessary to derive the evolution 
equations for higher-order correlation functions. However, 
the repetition of this treatment produces an infinite number of 
evolution equations and one cannot obtain a closed set of equations.   
This is the so-called {\it closure problem} for dynamics of 
statistical quantities. Note that the closure problem considered here 
is very close to the concept of BBGKY hierarchy, but slightly 
different in some sense. The BBGKY 
hierarchy arises from the many-body system characterized by the 
Loiuville equations and it also appears in the linear system. On the 
other hand, the origin of the closure problem essentially comes from the 
non-linearity of equation (\ref{eq:vec_fluid_eq}). Hence, 
to derive a closed set of moment equations, one must devise 
to introduce some truncation procedures by approximately treating 
the non-linear interaction in a self-consistent manner. 
The self-consistent truncation procedure is referred to as the closure theory 
(or closure approximation) in the statistical theory of turbulence 
and various closure theories have been so far exploited. 
In what follows, we especially consider the so-called 
{\it direct-interaction approximation} as one of the reliable closure 
theories.

Before closing this subsection, we define the propagator 
$G_{ab}(\bfk,\eta|\bfk',\eta')$, which will later play a key role 
in deriving a closed set of equations: 
\begin{equation}
G_{ab}(\bfk,\eta|\bfk',\eta')\equiv 
\frac{\delta \Phi_a(\bfk;\eta)}{\delta \Phi_b(\bfk';\eta')}, 
\end{equation}
where $\delta$ stands for a functional derivative. It represents 
the influence on $\Phi_a(\bfk;\eta)$ at time $\eta$ due to an infinitesimal
disturbance for $\Phi_b(\bfk';\eta')$ ($\eta\geq \eta'$). Taking a 
functional derivative of equation (\ref{eq:vec_fluid_eq}), we obtain 
the governing equation for the propagator as 
\begin{eqnarray}
&&\widehat{\mbox{\boldmath$\Lambda$}}_{ab}(\eta)\,
G_{bc}(\bfk,\eta|\bfk',\eta')=2\int\frac{d^3\bfk_1\,d^3\bfk_2}{(2\pi)^3}
\delta_D(\bfk-\bfk_1-\bfk_2)
\,\gamma_{apq}(\bfk_1,\bfk_2)\,
\nonumber\\
&&\quad\quad\quad\quad\quad\quad\quad\quad\quad\quad\quad\quad
\quad\quad\quad\quad\quad\quad\quad\quad\quad
\times\,\,
\Phi_p(\bfk_1;\eta)\,G_{qc}(\bfk_2,\eta|\bfk',\eta')
\label{eq:response_eq}
\end{eqnarray}
with the boundary condition: 
\begin{equation}
G_{ab}(\bfk,\eta'|\bfk',\eta')=  \delta_{ab}\,\,\delta_D(\bfk-\bfk').
\end{equation}

\section{Closure theory}
\label{sec:closure_theory}

\subsection{Direct-interaction approximation} 
\label{subsec:DI_approx}

In the subject of fluid mechanics, statistical characterization of 
turbulence for incompressible fluid flows is one of the major goals to 
understand the non-linear dynamics of Navier-Stokes equations. Among 
various attempts o construct a statistical theory of turbulence, there 
are approaches on systematic renormalized expansion 
\citep[e.g.,][]{K1959,W1961,L1973}. 
Direct-interaction approximation (DIA) is one of the 
best-known approximations and provides a simple truncation procedure 
\citep[][]{K1977,K1981,KG1997}.  DIA has several desirable properties 
such as the local energy conservation and the realizability of energy 
spectrum. Also, the agreement with numerical simulations of isotropic 
turbulence is excellent even at high Reynolds number. Although we are 
especially concerned with the dynamics of compressible and irrotational 
fluid flow, the non-linearity in the cosmic fluid system essentially come 
from the same advection terms as in the Navier-Stokes equations. In this 
respect, DIA is a promising method to give a quantitative prediction of 
matter power spectrum.

To derive a closed set of evolution equations in the DIA,  
one transparent and intuitive way is to decompose the true field $\Phi_a$ 
into the direct-interaction (DI) and the non-direct-interaction (NDI) parts. 
Let us consider particular Fourier modes $(\bfk,\, \bfp,\,\bfq)$ 
and especially focus on the time evolution of $\Phi_a$ for a specific 
Fourier mode $\bfk$. Through the non-linear term in the right-hand side of 
equation (\ref{eq:vec_fluid_eq}), 
the time evolution of $\Phi_a(\bfk;\eta)$ is determined by the infinite 
sum of three Fourier modes. 
We can formally decompose the quantity $\Phi_a(\bfk;\eta)$ into 
\begin{equation}
\Phi_a(\bfk;\eta)= \Phi_a^{\rm(NDI)}(\bfk;\eta|\bfp,\bfq)+ 
\Phi_a^{\rm(DI)}(\bfk;\eta|\bfp,\bfq).  
\end{equation}
In the above expression, 
the quantity  $\Phi_a^{\rm(DI)}$ denotes the DI field, whose 
time evolution is determined by the direct interaction with 
the particular Fourier modes, $\bfp$ and $\bfq$. On the other hand, 
the quantity $\Phi_a^{\rm(NDI)}$ is defined as a   
fictitious field without the direct interaction 
between three modes $\bfk$, $\bfp$ and $\bfq$.

In the system governed by dimensionless equation (\ref{eq:vec_fluid_eq}), 
the strength of the non-linearity is characterized by the number of 
interacting Fourier modes. In the non-linear regime, one naturally 
expects that NDI field $\Phi_a^{\rm(NDI)}$ plays 
the most dominant part in the time evolution of 
the quantity $\Phi_a(\bfk;\eta)$. 
Hence, in the DIA, we treat the DI field as a small perturbed quantity, 
relying on the assumption $\Phi_a^{\rm(NDI)}\gg\Phi_a^{\rm(DI)}$. 
In addition, we put the following assumptions: 
{\bf(i)} Gaussianity of the NDI field; 
{\bf(ii)} statistical independence among the modes without 
direct interaction; {\bf(iii)} statistical independence between 
the NDI field and the propagator. 
These assumptions basically come from the physical intuition for 
fully developed turbulence. In the presence of the infinite sum 
of the quadratic interaction, the fluid fields are expected to be 
nearly Gaussian, as naively indicated from the central-limit 
theorem. Also, it seems plausible that the initial correlation 
between different modes or quantities tends to be lost and the 
fluid quantities become statistically independent along the course 
of the non-linear interaction.  
Then, relying on these assumptions, 
we systematically expand the moment equations and the governing equation for 
propagator. Evaluating the ensemble average by using the formal 
solution of DI field in terms of the NDI fields, the  
closed set of equations for NDI fields can be finally derived 
\citep[][]{K1959,KG1997,GK1998}.

In the following, 
we shall derive a closed set of equations in an alternative route. 
As it has been shown by \citet{GK1998}, the resultant closure equations 
by DIA is identical with those 
obtained from the so-called {\it reversed expansion} 
procedure by introducing the fictitious 
parameter\footnote{In subject of turbulence, this fictitious parameter 
corresponds to the Reynolds number.} \citep[see also][]{L1973,K1977,K1981}.   
The reversed expansion procedure 
seems rather straightforward and the assumptions made in this procedure 
are relevant for the present cosmological situations.

\subsection{Derivation} 
\label{subsec:derivation}

Let us first introduce the fictitious parameter $\lambda$, which 
represents the strength of the non-linearity, and consider the 
weakly non-linear regime. We rewrite the field $\Phi_a$ and the 
propagator $G_{ab}$ with 
\begin{equation}
\Phi_a(\bfk;\eta)=\lambda \,\tdPhi_a(\bfk;\eta),\quad\quad
G_{ab}(\bfk,\eta|\bfk',\eta')=\lambda \,\tdG_{ab}(\bfk,\eta|\bfk',\eta').
\end{equation}
In terms of these, the basic equations 
(\ref{eq:vec_fluid_eq}) and (\ref{eq:response_eq}) respectively become 
\begin{eqnarray}
&&\widehat{\mbox{\boldmath$\Lambda$}}_{ab}(\eta)\,\, \tdPhi_b(\bfk;\eta)=
\lambda\,\int\frac{d^3\bfk_1\,d^3\bfk_2}{(2\pi)^3}
\delta_D(\bfk-\bfk_1-\bfk_2)\,\gamma_{abc}(\bfk_1,\bfk_2)\,
\tdPhi_b(\bfk_1;\eta)\,\tdPhi_c(\bfk_2;\eta),
\label{eq:basic_eq1}\\
&&\widehat{\mbox{\boldmath$\Lambda$}}_{ab}(\eta)\,\, 
\tdG_{bc}(\bfk,\eta|\bfk',\eta')=
2\,\,\lambda\,\int\frac{d^3\bfk_1\,d^3\bfk_2}{(2\pi)^3}
\delta_D(\bfk-\bfk_1-\bfk_2)\,\gamma_{apq}(\bfk_1,\bfk_2)\,
\nonumber\\
&&\quad\quad\quad\quad\quad\quad\quad\quad\quad\quad\quad\quad\quad
\quad\quad\quad\quad\quad\quad\quad\quad\quad
\times\,\,
\tdPhi_p(\bfk_1;\eta)\,\tdG_{qc}(\bfk_2,\eta|\bfk',\eta').
\label{eq:basic_eq2}
\end{eqnarray}

Our task is to derive the consistent closure equations from 
the moment equations (\ref{eq:moment_Pij}) and (\ref{eq:moment_Rij})  
with a help of the factitious parameter $\lambda$. 
To do this, we regard $\lambda$ as a small expansion parameter 
(i.e., $\lambda\ll1$) and put the following assumptions:  

\noindent
\underline{\it Assumptions}
\begin{description}
\item[(i)] The evolution of field $\tdPhi_a(\bfk;\eta)$ is started 
    from a tiny fluctuation and the development of non-linearity is mild. 
\item[(ii)] The field $\tdPhi_a(\bfk;\eta)$ is a homogeneous random field 
    and the statistical property of $\tdPhi_a(\bfk;\eta)$ is approximately 
    described by the Gaussian statistics. 
\item[(iii)] At the leading order, the field $\tdPhi_a$ and the propagator 
    $\tdG_{bc}$ are statistically independent from each other. 
\end{description}

Then, based on the perturbative calculation, we first 
evaluate the higher-order correlation terms in the moment equations 
(\ref{eq:moment_Pij}) and (\ref{eq:moment_Rij}). Inverting the 
perturbative expansion by a formal replacement of the perturbed quantities 
and setting the fictitious parameter to $\lambda=1$ at a final step, 
we obtain a closed set of evolution equations. The set 
of equations derived here may be regarded as the result of 
{\it renormalization} and/or {\it resummation} of the perturbative 
expansion, which will be applicable to the 
nonlinear regime of gravitational 
clustering beyond the standard perturbation theory.

\subsubsection{Naive perturbation} 
\label{subsubsec:naive_perturbation}

Based on the assumption {\bf (i)},  
let us evaluate the nonlinear terms by a perturbative expansion of the 
small parameter $\lambda$. 
To do this, we expand the quantities $\tdPhi_{a}$ and $\tdG_{ab}$ as:  
\begin{eqnarray}
\tdPhi_a(\bfk;\eta)&=&\tdPhi_a^{(0)}(\bfk;\eta)+
\lambda\,\, \tdPhi_a^{(1)}(\bfk;\eta)+\cdots,
\nonumber\\
\tdG_{ab}(\bfk,\eta|\bfk',\eta')&=&\tdG_{ab}^{(0)}(\bfk,\eta|\bfk',\eta')+
\lambda\,\, \tdG_{ab}^{(1)}(\bfk,\eta|\bfk',\eta)+\cdots. 
\label{eq:perturb_expansion}
\end{eqnarray}
The boundary condition for propagator in each order is
\begin{equation}
\tdG_{ab}^{(0)}(\bfk,\eta'|\bfk',\eta') =\delta_{ab}\,\delta_D(\bfk-\bfk'),
\quad\quad
\tdG_{ab}^{(1)}(\bfk,\eta'|\bfk',\eta') =0. 
\label{eq:bc_in_G}
\end{equation}
Also, the boundary condition for the first-order quantity  $\tdPhi_a^{(1)}$ 
at the initial time $\eta_0$ is 
\begin{equation}
\tdPhi_a^{(1)}(\bfk;\eta_0)=0. 
\label{eq:bc_in_u}
\end{equation}
Since the equations for zeroth-order quantities are source-free, 
no mode-mode coupling occurs. 
Thus, the zeroth-order propagator $\tdG_{ab}^{(0)}$ satisfying the 
boundary condition (\ref{eq:bc_in_G}) can be expressed in the following 
form: 
\begin{equation}
\tdG_{ab}^{(0)}(\bfk,\eta|\bfk',\eta')= 
\tdG_{ab}(\bfk|\eta,\eta')\,\,\delta_D(\bfk-\bfk').  
\label{eq:func_form_G0}
\end{equation}
Note that the function $\tdG_{ab}(\bfk|\eta,\eta')$ is 
a dimensionless quantity. 
Using this functional form, we formally write down the first-order 
solutions to the quantities $\tdPhi_i^{(1)}$ and $\tdG^{(1)}$. 
From (\ref{eq:basic_eq1}) and (\ref{eq:basic_eq2}), we obtain  
\begin{eqnarray}
\tdPhi_a^{(1)}(\bfk;\eta)&=&\int_{\eta_0}^{\eta} d\eta''\,\,
\tdG_{ac}(\bfk|\eta,\eta'')\,\,
\int\frac{d^3\bfk_1d^3\bfk_2}{(2\pi)^3}\,\delta_D(\bfk-\bfk_1-\bfk_2)
\nonumber\\
&&\quad\quad\quad\quad\quad\quad\times\,\,
\gamma_{cpq}(\bfk_1,\bfk_2)\,
\tdPhi_p^{(0)}(\bfk_1;\eta'')\,\tdPhi_q^{(0)}(\bfk_2;\eta''),
\label{eq:sol_u1}
\end{eqnarray}
\begin{eqnarray}
\tdG_{ab}^{(1)}(\bfk,\eta|\bfk',\eta')&=&2\,\int_{\eta'}^{\eta} d\eta''\,\,
\tdG_{ac}(\bfk|\eta,\eta'')\,\,
\int\frac{d^3\bfk_1d^3\bfk_2}{(2\pi)^3}\,\delta_D(\bfk-\bfk_1-\bfk_2)
\nonumber\\
&&\quad\quad\quad\quad\quad\quad\times\,\,
\gamma_{cpq}(\bfk_1,\bfk_2)\,
\tdPhi_p^{(0)}(\bfk_1;\eta'')\,\tdG_{qb}^{(0)}(\bfk_2,\eta''|\bfk',\eta')
\label{eq:sol_G1}
\end{eqnarray}
for $\eta>\eta'>\eta_0$. 

According to the assumption {\bf (ii)}, 
one may treat $\tdPhi_a^{(0)}$ as a Gaussian 
random variable.  Then, applying the perturbative expansion 
(\ref{eq:perturb_expansion}), we calculate the lowest-order 
non-vanishing contribution to 
the three-point correlation in equations (\ref{eq:moment_Pij}) 
and (\ref{eq:moment_Rij}). Let us first deal with the right-hand side of 
equation (\ref{eq:moment_Pij}). At the lowest-order contribution, 
the three-point correlation becomes 
\begin{eqnarray}
\Bigl\langle \tdPhi_a(\bfk;\eta)\,\tdPhi_p(\bfk_1;\eta)\,
\tdPhi_q(\bfk_2;\eta)\Bigr\rangle
&\simeq& \lambda\,\,
\Bigl\langle \tdPhi_a^{(0)}(\bfk;\eta)\,\tdPhi_p^{(0)}(\bfk_1;\eta)\,
\tdPhi_q^{(1)}(\bfk_2;\eta)\Bigr\rangle
\nonumber\\
&+&\lambda\,\,\Bigl\langle \tdPhi_a^{(0)}(\bfk;\eta)\,
\tdPhi_p^{(1)}(\bfk_1;\eta)\,
\tdPhi_q^{(0)}(\bfk_2;\eta)\Bigr\rangle
\nonumber\\
&+&\lambda\,\,\Bigl\langle \tdPhi_a^{(1)}(\bfk;\eta)\,
\tdPhi_p^{(0)}(\bfk_1;\eta)\,\tdPhi_q^{(0)}(\bfk_2;\eta)\Bigr\rangle,   
\end{eqnarray}
which are of the order of $\mathcal{O}(\lambda^1)$. Substituting the 
formal solution (\ref{eq:sol_u1}) into the above, we obtain
\begin{eqnarray}
&&\Bigl\langle 
\tdPhi_a(\bfk;\eta)\,\tdPhi_p(\bfk_1;\eta)\,\tdPhi_q(\bfk_2;\eta)\Bigr\rangle
\simeq\,\, \lambda\,
\int_{\eta_0}^{\eta} d\eta'\,\int \frac{d^3\bfp \,d^3\bfq}{(2\pi)^3}
\nonumber\\
&&\quad\quad
\times\Bigl[\,\,
\delta_D(\bfk_2-\bfp-\bfq)\,\tdG_{ql}(\bfk_2|\eta,\eta')\,
\gamma_{lrs}(\bfp,\bfq)
\,\,\Bigl\langle 
\tdPhi_a^{(0)}(\bfk;\eta)\,
\tdPhi_p^{(0)}(\bfk_1;\eta)\,
\tdPhi_r^{(0)}(\bfp;\eta')\,
\tdPhi_s^{(0)}(\bfq;\eta')
\Bigr\rangle
\nonumber\\
&&\quad\quad\quad
+\delta_D(\bfk_1-\bfp-\bfq)\,\tdG_{pl}(\bfk_1|\eta,\eta')\,
\gamma_{lrs}(\bfp,\bfq)\,\,\Bigl\langle 
\tdPhi_a^{(0)}(\bfk;\eta)\,
\tdPhi_r^{(0)}(\bfp;\eta')\,
\tdPhi_s^{(0)}(\bfq;\eta')\,
\tdPhi_q^{(0)}(\bfk_2;\eta)
\Bigr\rangle
\nonumber\\
&&\quad\quad\quad
+\delta_D(\bfk-\bfp-\bfq)\,\tdG_{al}(\bfk|\eta,\eta')\,
\gamma_{lrs}(\bfp,\bfq)\,\,\Bigl\langle 
\tdPhi_r^{(0)}(\bfp;\eta')\,
\tdPhi_s^{(0)}(\bfq;\eta')\,
\tdPhi_p^{(0)}(\bfk_1;\eta)\,
\tdPhi_q^{(0)}(\bfk_2;\eta)
\Bigr\rangle\,\,\Bigr].
\nonumber\\
&&\label{eq:three-pt}
\end{eqnarray}
The above expression is further reduced if we use the perturbative 
expression for the power spectra (\ref{eq:def_of_Pij_Rij}):
\begin{eqnarray}
&& P_{ab}(k;\eta)=\,\lambda^2\,\,\tdP_{ab}(k;\eta) 
 \simeq \lambda^2\,\,
\left\{ \tdP_{ab}^{(0)}(k;\eta) + \mathcal{O}(\lambda^2)\right\}, 
\nonumber\\
&& R_{ab}(k;\eta,\eta')=\,\lambda^2\,\,\tdR_{ab}(k;\eta,\eta')
\simeq \lambda^2\,\,
\left\{ \tdR_{ab}^{(0)}(k;\eta,\eta') + \mathcal{O}(\lambda^2) \right\}.
\label{eq:perturb_Pij_Rij}
\end{eqnarray}
The leading terms $\tdP_{ab}^{(0)}$ and $\tdR_{ab}^{(0)}$ are 
defined by 
\begin{eqnarray}
&&\Bigl\langle \tdPhi_a^{(0)}(\bfk;\eta)\,\tdPhi_b^{(0)}(\bfk';\eta) 
\Bigr\rangle = 
(2\pi)^3\,\delta_D(\bfk+\bfk')\,\tdP^{(0)}_{ab}(k;\eta), 
\nonumber\\
&&\Bigl\langle \tdPhi_a^{(0)}(\bfk;\eta)\,\tdPhi_b^{(0)}(\bfk';\eta') 
\Bigr\rangle = 
(2\pi)^3\,\delta_D(\bfk+\bfk')\,\tdR_{ab}^{(0)}(k;\eta,\eta'). 
\nonumber
\end{eqnarray}
Then, after some algebra, 
equation (\ref{eq:three-pt}) is finally reduced to the following form: 
\begin{eqnarray}
&\Bigl\langle \tdPhi_a(\bfk;\eta)\,\tdPhi_p(\bfk_1;\eta)\,
\tdPhi_q(\bfk_2;t)\Bigr\rangle \simeq& (2\pi)^3\,\,
\delta_D(\bfk+\bfk_1+\bfk_2)\,\,\lambda\,\,
F_{apq}^{(2)}(\bfk,\,\bfk_1,\,\bfk_2;\eta). 
\end{eqnarray}
The function $F_{apq}^{(2)}$ is explicitly written in terms of the 
two-time correlation $\tdR_{ab}^{(0)}$: 
\begin{eqnarray}
&&F_{apq}^{(2)}(\bfk,\bfk_1,\bfk_2;\eta)=2\,\,\int_{\eta_0}^{\eta} 
d\eta'\,\Bigl[\,
\tdG_{ql}(\bfk_2|\eta,\eta')\,\gamma_{lrs}(\bfk,\bfk_1)\,
\tdR_{ar}^{(0)}(k;\eta,\eta')\tdR_{ps}^{(0)}(k_1;\eta,\eta')
\nonumber\\
&&\,\,\quad\quad\quad\quad\quad\quad\quad\quad\quad\quad\quad+\,\,\,
\tdG_{pl}(\bfk_1|\eta,\eta')\,\gamma_{lrs}(\bfk,\bfk_2)\,
\tdR_{ar}^{(0)}(k;\eta,\eta')\,\tdR_{qs}^{(0)}(k_2;\eta,\eta')
\nonumber\\
&&\,\,\quad\quad\quad\quad\quad\quad\quad\quad\quad\quad\quad+\,\,\,
\tdG_{al}(\bfk|\eta,\eta')\,\gamma_{lrs}(\bfk_1,\bfk_2)\,
\tdR_{pr}^{(0)}(k_1;\eta,\eta')\,\tdR_{qs}^{(0)}(k_2;\eta,\eta')
\,\Bigr]. 
\label{eq:kernel_F}
\end{eqnarray}
In deriving equation (\ref{eq:kernel_F}), 
we have used the symmetric properties of the 
vertex function, i.e., 
$\gamma_{abc}(\bfk_1,\bfk_2)=\gamma_{acb}(\bfk_2,\bfk_1)$ and 
$\gamma_{abc}(\bfk,-\bfk)=0$.

In similar manner, we perturbatively evaluate the three-point correlation  
in equation (\ref{eq:moment_Rij}). The resultant expression becomes  
\begin{eqnarray}
\Bigl\langle \tdPhi_c(\bfk';\eta')\,\tdPhi_p(\bfk_1;\eta)\,
\tdPhi_q(\bfk_2;\eta)\Bigr\rangle 
&\simeq& (2\pi)^3\,\,\delta_D(\bfk'+\bfk_1+\bfk_2)\,\,\lambda\,\,K_{cpq}^{(2)}
(\bfk',\bfk_1,\bfk_2;\eta,\eta'),
\nonumber\\
&&\quad\quad\quad\quad\quad\quad\quad\quad\quad\quad\quad\quad\quad
\quad\quad\quad\,\,\,\,(\eta>\eta')
\end{eqnarray}
with the function $K_{bpq}^{(2)}$ being 
\begin{eqnarray}
&&K_{cpq}^{(2)}(\bfk',\bfk_1,\bfk_2;\eta,\eta')
\nonumber\\
&&
=\,\,2\,\,\int_{\eta_0}^{\eta} d\eta''\,
\Bigl\{ \tdR_{cr}^{(0)}(k';\eta',\eta'')\Theta(\eta'-\eta'')+
\tdR_{rc}^{(0)}(k';\eta'',\eta')\Theta(\eta''-\eta')\Bigr\}
\nonumber\\
&&\quad\times\,\,\Bigl[\,
\tdG_{ql}(\bfk_2|\eta,\eta'')\,\gamma_{lrs}(\bfk',\bfk_1)\,
\tdR_{ps}^{(0)}(k_1;\eta,\eta'') + \tdG_{pl}(\bfk_1|\eta,\eta'')
\,\gamma_{lrs}(\bfk',\bfk_2)\,\tdR_{qs}^{(0)}(k_2;\eta,\eta'')\,\Bigr]
\nonumber\\
&&\,+ \,\,2\,\,\int_{\eta_0}^{\eta'} d\eta''\,
\tdG_{cl}(\bfk'|\eta',\eta'')\,\gamma_{lrs}
(\bfk_1,\bfk_2)\,R_{pr}^{(0)}(k_1;\eta,\eta'')\,
\tdR_{qs}^{(0)}(k_2;\eta,\eta''). 
\label{eq:kernel_K}
\end{eqnarray}
Here, the function $\Theta(t)$ is the Heaviside step function.

Now, summing up the perturbative expressions of three-point correlations,  
the moment equations (\ref{eq:moment_Pij}) and (\ref{eq:moment_Rij}) 
respectively become 
\begin{eqnarray}
\widehat{\mbox{\boldmath$\Sigma$}}_{abcd}(\eta)\,
\tdP_{cd}(k;\eta)&\simeq& \lambda^2
\int \frac{d^3\bfq}{(2\pi)^3}\,
\Bigl[\,\gamma_{bpq}(\bfq,-\bfk-\bfq)\,F_{apq}^{(2)}
(\bfk,\bfq,-\bfk-\bfq;\,\eta)
\nonumber \\
&&\quad\quad\quad\quad\quad +\,
\gamma_{apq}(\bfq,\bfk-\bfq)\,F_{bpq}^{(2)}(-\bfk,\bfq,\bfk-\bfq;\,\eta)\,
\Bigr]\, +\, \mathcal{O}(\lambda^4),
\label{eq:perturb_rhs1}
\\
\widehat{\mbox{\boldmath$\Lambda$}}_{ab}(\eta)\,
\tdR_{bc}(k;\eta,\eta')&\simeq& \lambda^2
\int \frac{d^3\bfq}{(2\pi)^3}\,
\,\gamma_{apq}(\bfq,\bfk-\bfq)\,K_{cpq}^{(2)}
(-\bfk,\bfq,\bfk-\bfq;\,\eta,\eta')\, + \,\mathcal{O}(\lambda^4).
\label{eq:perturb_rhs2}
\end{eqnarray}

\subsubsection{Non-linear propagator}
\label{subsubsec:propagator}

Next, we evaluate the higher-order terms in the 
governing equation for propagator (\ref{eq:basic_eq2}). 
To do this, we first notice that the propagator $\tdG_{ab}$ is no 
longer deterministic. Because of the interaction with the 
random field $\tdPhi_a$, the time evolution of $\tdG_{ab}$ also exhibits 
stochastic nature. We thus treat the 
equation (\ref{eq:basic_eq2}) statistically: 
\begin{eqnarray}
&\widehat{\mbox{\boldmath$\Lambda$}}_{ab}(\eta)
\Bigl\langle \tdG_{bc}(\bfk,\eta|\bfk',\eta')\Bigr\rangle&=
2\,\lambda\,\int\frac{d^3\bfk_1\,d^3\bfk_2}{(2\pi)^3}
\delta_D(\bfk-\bfk_1-\bfk_2)\,\gamma_{apq}(\bfk_1,\bfk_2)\,
\nonumber\\
&&\quad\quad\quad\quad\quad\quad\quad\quad\quad\quad
\times\,\,
\Bigl\langle \tdPhi_p(\bfk_1;\eta)\,\tdG_{qc}(\bfk_2,\eta|\bfk',\eta') 
\Bigr\rangle.
\label{eq:ensemble_response}
\end{eqnarray}
The higher-order term in the right-hand side of equation 
(\ref{eq:ensemble_response}) is perturbatively evaluated as  
\begin{eqnarray}
\Bigl\langle \tdPhi_p(\bfk_1;\eta)\,\tdG_{qc}(\bfk_2,\eta|\bfk',\eta') 
\Bigr\rangle
&\simeq&
\Bigl\langle  
\tdPhi_p^{(0)}(\bfk_1;\eta)\,\tdG_{qc}^{(0)}(\bfk_2,\eta|\bfk',\eta') 
\Bigr\rangle
\nonumber\\
&+& \lambda\,\,
\Bigl\langle  
\tdPhi_p^{(1)}(\bfk_1;\eta)\,\tdG_{qc}^{(0)}(\bfk_2,\eta|\bfk',\eta') 
\Bigr\rangle
\nonumber\\
&+& \lambda\,\,
\Bigl\langle  
\tdPhi_p^{(0)}(\bfk_1;\eta)\,\tdG_{qc}^{(1)}(\bfk_2,\eta|\bfk',\eta') 
\Bigr\rangle. 
\label{eq:three_terms}
\end{eqnarray}
In the above equation, the first two terms at the right-hand side 
become vanishing because of the assumption {\bf(iii)}. 
The only non-vanishing contribution comes from the last term,  
which can be recast as 
\begin{eqnarray}
\Bigl\langle  
\tdPhi_p^{(0)}(\bfk_1;\eta)\,\tdG_{qc}^{(1)}(\bfk_2,\eta|\bfk',\eta') 
\Bigr\rangle &= & 2 
\int_{\eta'}^{\eta} d\eta'' \tdG_{ql}(\bfk_2|\eta,\eta'')\,\,
\int \frac{d^3\bfp d^3\bfq}{(2\pi)^3}\,\delta_D(\bfk_2-\bfp-\bfq)
\nonumber\\
&&\quad \times \,\gamma_{lrs}(\bfp,\bfq)\,
\Bigl\langle \tdPhi^{(0)}_p(\bfk_1;\eta)\tdPhi^{(0)}_r(\bfp;\eta'')
\Bigr\rangle 
\,\tdG^{(0)}_{sc}(\bfq,\eta''|\bfk',\eta') 
\nonumber\\
&= & 2 \,\,\delta_D(\bfk'-\bfk_1-\bfk_2)\,\,\int_{\eta'}^{\eta} d\eta''\,\,
\tdG_{ql}(\bfk_2|\eta,\eta'')
\nonumber\\
&&\quad 
\times \,\gamma_{lrs}(-\bfk_1,\bfk')\,\,\tdR_{pr}^{(0)}(k_1;\eta,\eta'')\,\,
\tdG_{sc}(\bfk'|\eta,\eta'').
\end{eqnarray}
In the last equality, we have used equation (\ref{eq:func_form_G0}) and 
the definition of $\tdR_{ab}^{(0)}$ (see Eq.[\ref{eq:perturb_Pij_Rij}] below). 
Hence, the perturbative evaluation of equation (\ref{eq:basic_eq2}) 
becomes 
\begin{eqnarray}
\widehat{\mbox{\boldmath$\Lambda$}}_{ab}(\eta)\,
\Bigl\langle \tdG_{bc}(\bfk,\eta|\bfk',\eta') \Bigr\rangle
&\simeq& \,\,4\,\lambda^2\,\,\delta_D(\bfk-\bfk')
\int_{\eta'}^{\eta} d\eta'' \int \frac{d^3\bfq}{(2\pi)^3}\,
\,\gamma_{apq}(\bfq,\bfk-\bfq)\,\gamma_{lrs}(-\bfq,\bfk') 
\nonumber\\
&&\quad
\times\,\,\tdG_{ql}(|\bfk-\bfq||\eta,\eta'')\,
\tdR_{pr}^{(0)}(q|;\eta,\eta'')\,\tdG_{sc}(\bfk'|\eta'',\eta')
\, + \,\mathcal{O}(\lambda^4).
\nonumber\\
\label{eq:perturb_rhs3}
\end{eqnarray}
up to the contribution of $\mathcal{O}(\lambda^2)$.

\subsubsection{Reversed expansion}
\label{subsubsec:reversion}

We are in position to employ the procedure of the so-called {\it reversion} 
to rewrite  the perturbative expressions 
(\ref{eq:perturb_rhs1}), (\ref{eq:perturb_rhs2}) and 
(\ref{eq:perturb_rhs3}).

Recall from the perturbative expansion (\ref{eq:perturb_Pij_Rij}) 
that the $\mathcal{O}(\lambda^4)$ and the higher-order terms of the 
power spectra can be expressed in terms of the $\tdP_{ab}^{(0)}$ and 
$\tdR_{ab}^{(0)}$ in principle,   
since the formal solution of higher-order quantity $\tdPhi_a^{(n)}$ is always
written in terms of $\tdPhi_a^{(0)}$ with a help of propagator 
of zeroth-order, $\tdG_{ab}^{(0)}$. 
Also, the ensemble average of 
the higher-order propagator is formally expressed in terms of 
the zeroth-order quantities, $\tdG_{ab}^{(0)}$ and $\tdR_{ab}^{(0)}$.  
We then regard the expansion (\ref{eq:perturb_Pij_Rij}) as 
equations for $\tdP_{ab}$ and $\tdR_{ab}$, the solutions of which are 
written in powers of $\lambda$ as 
\begin{equation}
\tdP_{ab}^{(0)}(k;\eta)=\tdP_{ab}(k;\eta)+ \mathcal{O}(\lambda^2), 
\quad \quad 
\tdR_{ab}^{(0)}(k;\eta,\eta')=\tdR_{ab}(k;\eta,\eta')+ \mathcal{O}(\lambda^2). 
\end{equation}
Similarly, we may write 
\begin{equation}
\delta_D(\bfk-\bfk')\,\,\tdG_{ab}(\bfk|\eta,\eta') \,=
\Bigl\langle \tdG_{ab}(\bfk,\eta|\bfk',\eta') \Bigr\rangle 
+\,\mathcal{O}(\lambda^2). 
\end{equation}
This procedure is called the reversion and it corresponds to 
the resummation of the perturbation series. Thus, 
at the leading order, 
equations (\ref{eq:perturb_rhs1}), (\ref{eq:perturb_rhs2}) 
and (\ref{eq:perturb_rhs3}) are written 
in terms of the true field variables $\tdP_{ab}$, $\tdR_{ab}$ 
and $\tdG_{ab}$.

Now, we do not necessarily treat $\lambda$ as the small parameter. 
This is just a book-keeping parameter and we finally set it to unity. 
Dropping the tilde over the quantities $P_{ab}$, $R_{ab}$ 
and $G_{ab}$,  we at last reach the closure equations:  
\begin{eqnarray}
&&\widehat{\mbox{\boldmath$\Sigma$}}_{abcd}(\eta)\,P_{cd}(k;\eta)=
\,\,
\int \frac{d^3\bfq}{(2\pi)^3}\,
\Bigl[\,\gamma_{bpq}(\bfq,-\bfk-\bfq)\,F_{apq}
(\bfk,\bfq,-\bfk-\bfq;\,\eta)
\nonumber \\
&&\quad\quad\quad\quad\quad\quad\quad\quad\quad\quad\quad\quad\quad
\quad\quad\quad +
\gamma_{apq}(\bfq,\bfk-\bfq)\,F_{bpq}(-\bfk,\bfq,\bfk-\bfq;\,\eta)\,
\Bigr],
\label{eq:DIA_eq1}\\
&&\widehat{\mbox{\boldmath$\Lambda$}}_{ab}(\eta)\,R_{bc}(k;\eta,\eta')=
\,\,
\int \frac{d^3\bfq}{(2\pi)^3}\,
\,\gamma_{apq}(\bfq,\bfk-\bfq)\,K_{cpq}
(-\bfk,\bfq,\bfk-\bfq;\,\eta,\eta'), 
\label{eq:DIA_eq2}\\
&&\widehat{\mbox{\boldmath$\Lambda$}}_{ab}(\eta)\,
 G_{bc}(\bfk|\eta,\eta')=\,\,4\,
\int_{\eta'}^{\eta} d\eta'' \int \frac{d^3\bfq}{(2\pi)^3}\,
\,\gamma_{apq}(\bfq,\bfk-\bfq)\,\gamma_{lrs}(-\bfq,\bfk) 
\nonumber\\
&&\quad\quad\quad\quad\quad\quad\quad\quad\quad\quad
\quad\quad\quad \times G_{ql}(|\bfk-\bfq||\eta,\eta'')\,
R_{pr}(q;\eta,\eta'')\,G_{sc}(\bfk|\eta'',\eta').
\label{eq:DIA_eq3}
\end{eqnarray}
The explicit expressions for the kernels $F_{apq}$ and $K_{cpq}$ are 
summarized as 
\begin{eqnarray}
&&F_{apq}(\bfk,\bfk_1,\bfk_2;\eta)=2\int_{\eta_0}^{\eta} d\eta''\,\Bigl[\,
2\,\,G_{ql}(k_2|\eta,\eta'')\,\gamma_{lrs}(\bfk,\bfk_1)\,
R_{ar}(k;\eta,\eta'')R_{ps}(k_1;\eta,\eta'')
\nonumber\\
&&\quad\quad\quad\quad\quad\quad\quad\quad\quad\quad+\,\,\,
G_{al}(k|\eta,\eta'')\,\gamma_{lrs}(\bfk_1,\bfk_2)\,
R_{pr}(k_1;\eta,\eta'')\,R_{qs}(k_2;\eta,\eta'')
\,\Bigr], 
\label{eq:kernel_F2}\\
 &&K_{cpq}(\bfk',\bfk_1,\bfk_2;\eta,\eta')
 \nonumber\\
 &&\quad\quad\quad
 =\,\,4\int_{\eta_0}^{\eta} d\eta''\,
 \,G_{ql}(k_2|\eta,\eta'')\,\gamma_{lrs}(\bfk',\bfk_1)\,
 R_{ps}(k_1;\eta,\eta'') 
 \nonumber\\
 &&\quad\quad\quad\quad\quad\quad\quad\quad\times\,\,
 \Bigl\{ R_{cr}(k';\eta',\eta'')\Theta(\eta'-\eta'')+
 R_{rc}(k';\eta'',\eta')\Theta(\eta''-\eta')\Bigr\}
 \nonumber\\
 &&\quad\quad\quad\,+ \,\,2\int_{\eta_0}^{\eta'} d\eta''\,
 G_{cl}(k'|\eta',\eta'')\,\gamma_{lrs}(\bfk_1,\bfk_2)\,
R_{pr}(k_1;\eta,\eta'')\, R_{qs}(k_2;\eta,\eta''),   
\label{eq:kernel_K2}
\end{eqnarray}
where we have used the fact that the functions $F_{apq}$ and $K_{cpq}$ 
always appear as the product of $\gamma_{bpq}F_{apq}$ and 
$\gamma_{apq}K_{cpq}$.

\section{Properties of closure equations}
\label{sec:properties}

In this section, we discuss some important properties of the 
closure equations derived in previous section.

\subsection{Recovery of one-loop perturbations}
\label{subsec:recovery}

The closure equations in previous section have been derived 
in somewhat non-trivial manner by the reversed expansion procedure.  
While we employ the perturbative approach when evaluating the 
moment equations, the final governing equations for matter power spectrum 
become the non-linear coupled system and it seems unclear 
whether the power spectra calculated from the closure equations consistently 
recover the results of the standard perturbation theory at some level. 
In this respect, the recovery of the perturbation calculation may be 
a fast important check for the usefulness of the closure approximation.

In the standard treatment of the perturbation theory,  
the field $\Phi_a$ is assumed to be a small perturbed quantity and 
is expanded as 
\begin{equation}
\Phi_a=\Phi_a^{(1)}+\Phi_a^{(2)}+\Phi_a^{(3)}+\cdots.
\end{equation}
Substituting the above expansion into the evolution equation 
(\ref{eq:vec_fluid_eq}), we systematically derive the perturbation 
equations and through the order-by-order treatment,  
the solutions for higher-order quantities are expressed 
in terms of the linear-order quantity $\Phi_a^{(1)}$. 
Further assuming the Gaussianity of the linear-order quantity 
$\Phi_a^{(1)}$, the power spectra can be summarized as 
\begin{equation}
P_{ab}(k)=P_{ab}^{(11)}(k)+\,\,
\left\{P_{ab}^{(22)}(k)+P_{ab}^{(13)}(k)\right\}+\cdots, 
\end{equation}
where the first term in the right-hand-side of equation is the 
linear power spectra and the quantities in the curly bracket
is the so-called one-loop corrections to the power spectra, given by  
\begin{eqnarray}
\Bigl\langle\Phi_a^{(1)}(\bfk;\eta)\Phi_b^{(1)}(\bfk';\eta)\Bigr\rangle &=& 
(2\pi)^3\,\delta_D(\bfk+\bfk')\,\,P_{ab}^{(11)}(k;\eta),
\nonumber\\
\Bigl\langle\Phi_a^{(2)}(\bfk;\eta)\Phi_b^{(2)}(\bfk';\eta)\Bigr\rangle &=& 
(2\pi)^3\,\delta_D(\bfk+\bfk')\,\,P_{ab}^{(22)}(k;\eta),
\nonumber\\
\Bigl\langle\Phi_a^{(1)}(\bfk;\eta)\Phi_b^{(3)}(\bfk';\eta)+
\Phi_a^{(3)}(\bfk;\eta)\Phi_b^{(1)}(\bfk';\eta)\Bigr\rangle &=& 
(2\pi)^3\,\delta_D(\bfk+\bfk')\,\,P_{ab}^{(13)}(k;\eta).
\end{eqnarray}

In Appendix \ref{appendix:one-loop_PT}, we show that the linear plus 
one-loop power spectra satisfy the following evolution equations: 
\begin{eqnarray}
&&\widehat{\mathbf{\Sigma}}_{abcd}(\eta)\,
\left\{
P_{cd}^{(11)}(k)+ P_{cd}^{(22)}(k)+P_{cd}^{(13)}(k)
\right\}
\nonumber\\
&&\quad\quad\quad\quad=\int \frac{d^3\bfk_1d^3\bfk_2}{(2\pi)^3}
\Bigl\{
\delta_D(\bfk+\bfk_1+\bfk_2)\,\gamma_{bpq}(\bfk_1,\bfk_2)\,
F_{apq}(\bfk,\bfk_1,\bfk_2;\eta)\Bigr.
\nonumber\\
&&\quad\quad\quad\quad\quad\quad\quad\quad\quad\quad~~
+\delta_D(\bfk-\bfk_1-\bfk_2)\,\gamma_{apq}(\bfk_1,\bfk_2)\,
F_{bpq}(-\bfk,\bfk_1,\bfk_2;\eta)\Bigr.
\Bigr\}, 
\label{eq:recovery_1loopPT}
\end{eqnarray}
where the function $F_{apq}$ exactly coincides with 
the definition (\ref{eq:kernel_F2}), with the replacement of 
the non-linear propagator and power spectra, $G_{ab}$ and $R_{ab}$, 
with those of the linear counterparts, $g_{ab}$ and $R_{ab}^{(11)}$:  
\begin{eqnarray}
\widehat{\mbox{\boldmath$\Lambda$}}_{ab}(\eta)\,\,g_{bc}(\eta,\eta')
=\delta_{ac}\,\delta_D(\eta-\eta') 
\label{eq:eq_linear_propagator}
\end{eqnarray}
and
\begin{eqnarray}
\Bigl\langle\Phi_a^{(1)}(\bfk;\eta)\Phi_b^{(1)}(\bfk';\eta')\Bigr\rangle 
= (2\pi)^3\,\delta_D(\bfk+\bfk')\,\,R_{ab}^{(11)}(k;\eta,\eta'),\quad
(\eta\geq \eta').
\end{eqnarray}

Thus, in the weakly non-linear regime, the closure approximation 
faithfully recover the one-loop results of the standard perturbation theory. 
This will be manifestly apparent in next section by calculating the 
power spectra. A great emphasis is that among several non-perturbative 
approaches, the closure equations as non-linear coupled system 
also have the potential to go beyond the perturbation theory. In fact, 
our closure theory is basically equivalent to the the 2PI effective 
action approach by \citet{V2007a} and 
the one-loop level of the RPT by 
\citet{CS2006a}. This will be explicitly shown in next subsection 
when we obtain the exact integral expressions. 


\subsection{Exact integral solutions}
\label{subsec:integral_solutions}

The closure equations derived before seem rather complicated and 
analytically intractable because of its non-linearity and non-locality. 
In practice, a sophisticated numerical treatment is required to 
get the exact solutions for closure system. Note, however, 
that the closure equations possess 
the {\it exact} integral expressions for the power spectra 
$P_{ab}$ and $R_{ab}$, which are formal solutions of the closure 
equations:  
\begin{eqnarray}
P_{ab}(k;\eta)&=&G_{ac}(k|\eta,\eta_0)\,
G_{bd}(k|\eta,\eta_0)\,P_{cd}(k;\eta_0)
\nonumber\\
&&\quad\quad+\,\int_{\eta_0}^{\eta}d\eta_1 \int_{\eta_0}^{\eta}d\eta_2\,
G_{ac}(k|\eta,\eta_1)G_{bd}(k|\eta,\eta_2)\Phi_{cd}(k;\eta_2,\eta_1).
\label{eq:integral_1}
\\
R_{ab}(k;\eta,\eta')&=&G_{ac}(k|\eta,\eta_0)\,
G_{bd}(k|\eta',\eta_0)\,P_{cd}(k;\eta_0)
\nonumber\\
&&\quad\quad+\,\int_{\eta_0}^{\eta}d\eta_1 \int_{\eta_0}^{\eta'}d\eta_2\,
G_{ac}(k|\eta,\eta_1)G_{bd}(k|\eta',\eta_2)\Phi_{cd}(k;\eta_2,\eta_1),  
\label{eq:integral_2}
\end{eqnarray}
The above expressions contain the function $\Phi(k;\eta_1,\eta_2)$,  
which represents the non-linear mode-coupling between different Fourier 
modes, given by 
\begin{eqnarray}
&&\Phi_{ab}(k;\eta_1,\eta_2)=2\,\int\frac{d^3\bfq}{(2\pi)^3}\,
\gamma_{ars}(\bfq,\bfk-\bfq)\,\gamma_{bpq}(\bfq,\bfk-\bfq)
\nonumber\\
&&\quad\quad\quad\quad\quad\times\,
\Bigl\{
R_{pr}(q;\eta_1,\eta_2)R_{qs}(|\bfk-\bfq|;\eta_1,\eta_2)\,
\Theta(\eta_1-\eta_2) 
\nonumber\\
&&\quad\quad\quad\quad\quad\quad\quad\quad\quad\quad\quad\quad\quad\quad+ 
R_{rp}(q;\eta_2,\eta_1)R_{sq}(|\bfk-\bfq|;\eta_2,\eta_1)\,
\Theta(\eta_2-\eta_1)
\Bigr\}. 
\label{eq:mode_coupling}
\end{eqnarray}
Note that the mode-coupling function $\Phi$ 
possesses the following symmetry: 
$\Phi_{ab}(k;\eta_1,\eta_2)=\Phi_{ba}(k;\eta_2,\eta_1)$. 
In Appendix \ref{appendix:integral_sol}, the integral expressions 
(\ref{eq:integral_1}) and (\ref{eq:integral_2}) 
are indeed compatible with the closure equations if the 
mode-coupling function $\Phi$ is given by equation 
(\ref{eq:mode_coupling}).

The integral expressions given above have been also derived 
based on the RPT 
by \citet{CS2006a} and/or through the path-integral formulation 
by \citet{MP2007} \citep[see also][]{V2007a, CS2007}, although 
their derivations are quite formal. In contrast to their formal 
expressions, our integral solutions have the explicit functional 
dependence of 
the mode-coupling function $\Phi_{ab}$ on the power spectra $R_{ab}$ and 
the propagators $G_{ab}$. In the language of
the RPT, this 
corresponds to the renormalized expressions for the mode-coupling power 
up to the one-loop order\footnote{Correctly speaking, 
closure approximation or DIA 
drops all the corrections arising from the vertex renormalization 
\citep{W1961}.}. 
In this respect, the closure equations is a 
non-perturbative description of the power spectra going 
beyond the perturbation theory, and 
have an ability to predict the matter power spectra accurately, 
the result of which will be
comparable to the one-loop results from RPT
or the path-integral approach. Note, however, that our closure system 
has time evolution of the non-linear propagator, 
whose governing equation has been also derived 
by the self-consistent truncation of the higher-order corrections. 
On the other hand, in the RPT, 
no such truncation is considered in deriving the integral expressions. 
This may cause a major difference in the prediction of matter power 
spectra, which we will address in detail 
in next section.

\section{Analytical treatment of non-linear power spectrum}
\label{sec:analytic_treatment}

In this section, we compute the power spectrum of mass density 
fluctuation from the closure equations.  
Based on the exact integral expressions, 
we employ the Born approximation to obtain the analytic expressions 
for power spectrum. Further, the approximate expressions for the 
non-linear propagator is obtained by matching the two asymptotic behaviors.  
Combining these results, we evaluate the power spectrum of mass fluctuations 
and compare it with the one obtained 
from the RPT \citep{CS2007}. 
In what follows, the following cosmological parameters are adopted to 
compute the power spectra :  
$\Omega_{\rm m,0}=0.27$, $\Omega_{\rm b,0}=0.043$, $\Omega_{w,0}=0.73$, $w=-1$, 
$h=0.7$, $\sigma_8=0.8$, and $n_s=1$.

\subsection{Born approximation}
\label{subsec:Born_approx}

In practice, numerical treatment to directly solve the closure 
equations would be essential for an accurate evaluation of non-linear 
power spectrum. Nevertheless, analytical evaluation of the power spectrum 
is instructive and very helpful to understand the behavior of 
the non-linear corrections incorporated into the closure system. 
\citet{CS2007} recently applied the first-order Born approximation to 
the exact integral expressions for the power spectra and derived the 
analytic expressions correctly up to the two-loop order. 
In the following, adopting the same approximation, we will analytically 
evaluate the power spectrum.

Here, the term, Born approximation, means the iterative approximation 
scheme to evaluate the integral equations like 
(\ref{eq:integral_1}) and (\ref{eq:integral_2}).
Applying the Born approximation to the integral solutions, 
the power spectra $P_{ab}$ and $R_{ab}$ 
are first evaluated by substituting the 
linear-order quantities into the right-hand side 
of the expressions (\ref{eq:integral_1}) and (\ref{eq:integral_2}).   
This is the first-order Born approximation and  
it can be further 
improved by repeating the iterative substitution of 
the leading-order solutions into the right-hand side of the 
integral solutions. This treatment can be accurate in principle and 
the higher-order corrections become negligible 
as long as the contribution from the mode-coupling function is small,  
compared to the first term in right-hand side of integral equations.

In next subsection, we will present the approximate 
expression for the non-linear propagator, 
$G_{ab}^{\rm approx}$. Here, assuming the analytic form of 
$G_{ab}$, we derive analytical expressions for the 
quantity $P_{ab}(k;\eta)$ 
by substituting the iterative solutions of the different-time power 
spectra $R_{ab}(k;\eta,\eta')$. Let us denote the linear power spectra 
given at initial time by $P_{ab}^{\rm lin}(k;\eta_0)$. 
For a sufficiently small value of $\eta_0$, the late-time evolution 
is dominated by the growing-mode solution. We thus put
\begin{equation}
P_{ab}^{\rm lin}(k;\eta_0)=e^{2\eta_0} P_0(k) 
\left(
\begin{array}{cc}
1 & 1 \\ 1 & 1
\end{array}
\right) 
\end{equation}
with the quantity $P_0(k)$ being the linearly extrapolated spectrum 
at the present time. Then, from equation (\ref{eq:integral_2}), 
different-time power spectra are iteratively evaluated as follows: 
\begin{eqnarray}
R_{ab}(k;\eta,\eta')=R_{ab}^{\rm (I)}(k;\eta,\eta')+ 
R_{ab}^{\rm (II)}(k;\eta,\eta') + \cdots\,\,;
\nonumber
\end{eqnarray}
\begin{eqnarray}
R_{ab}^{\rm (I)}(k;\eta,\eta')&=&
\widetilde{G}_{a}(k|\eta,\eta_0)
\widetilde{G}_{b}(k|\eta',\eta_0)
e^{2\eta_0}\,P_0(k;\eta_0), 
\nonumber\\
R_{ab}^{\rm (II)}(k;\eta,\eta') &=& 2\int\frac{d^3\bfq}{(2\pi)^3}\,
I_{a}(\bfk,\,\bfq;\eta,\,\eta_0)\,
I_{b}(\bfk,\,\bfq;\eta',\,\eta_0)
\,\,e^{4\eta_0} P_0(q)\, P_0(|\bfk-\bfq|).  
\nonumber
\end{eqnarray}
with $\widetilde{G}_a\equiv G_{a1}+G_{a2}$. Substituting 
the iterative solutions of $R_{ab}$ into the integral 
expression (\ref{eq:integral_1}), one obtains 
\begin{eqnarray}
P_{ab}(k;\eta)=P_{ab}^{\rm(I)}(k;\eta) + 
P_{ab}^{\rm(II)}(k;\eta)+P_{ab}^{\rm(III)}(k;\eta)+\cdots   ;
\label{eq:Pk_Born}
\end{eqnarray}
\begin{eqnarray}
P_{ab}^{\rm(I)}(k;\eta)&=&\widetilde{G}_{a}(k|\eta,\eta_0) 
\widetilde{G}_{b}(k|\eta,\eta_0) e^{2\eta_0} P_0(k), 
\nonumber\\
P_{ab}^{\rm(II)}(k;\eta)&=& 2\int\frac{d^3\bfq}{(2\pi)^3}\,
I_{a}(\bfk,\,\bfq;\eta,\,\eta_0)\,
I_{b}(\bfk,\,\bfq;\eta,\,\eta_0)
\,\,e^{4\eta_0} P_0(q)\, P_0(|\bfk-\bfq|), 
\nonumber\\
P_{ab}^{\rm (III)}(k;\eta) &=&
8\int\frac{d^3\bfp}{(2\pi)^3}\int\frac{d^3\bfq}{(2\pi)^3} 
J_a(\bfk,\bfp,\bfq;\eta,\,\eta_0)J_b(\bfk,\,\bfp,\,\bfq;\eta,\,\eta_0)
\nonumber\\
&&\quad\quad\quad\quad\quad\times\,\,e^{6\eta_0} \,\,
P_0(|\bfk-\bfp|)P_0(q)P_0(|\bfp-\bfq|).
\nonumber
\end{eqnarray}
Here, the source functions $I_a$ and $J_a$ are respectively given by 
\begin{eqnarray}
&&I_{a}(\bfk,\,\bfq;\eta,\,\eta_0)=\int_{\eta_0}^{\eta}d\eta'\,
G_{al}(k|\eta,\eta')\,\gamma_{lrs}(\bfq,\bfk-\bfq)
\,\widetilde{G}_{r}(q|\eta',\eta_0)\,
\widetilde{G}_{s}(|\bfk-\bfq||\eta',\eta_0),
\nonumber\\
&&J_{a}(\bfk,\,\bfp,\,\bfq;\eta,\,\eta_0)=
\int_{\eta_0}^{\eta}d\eta_1\,\int_{\eta_0}^{\eta}d\eta_2
G_{al}(k|\eta,\eta_1)\,\gamma_{lrs}(\bfp,\bfk-\bfp)\,G_{rc}(p|\eta_1,\eta_2)
\nonumber\\
&&\quad\quad\quad\quad\quad\quad\quad\quad\quad
\times\,\,\,\gamma_{cpq}(\bfq,\bfp-\bfq)\,
\widetilde{G}_{p}(q|\eta_2,\eta_0) 
\widetilde{G}_{q}(|\bfp-\bfq||\eta_2,\eta_0) 
\widetilde{G}_{s}(|\bfk-\bfp||\eta_1,\eta_0).  
\nonumber
\end{eqnarray}

Compared to 
the results given by \cite{CS2007}, we find that the first and second 
terms in equation (\ref{eq:Pk_Born}), 
$P_{ab}^{\rm(I)}$ and $P_{ab}^{\rm(II)}$, 
exactly coincide with their expressions, $G^2P_0$ and 
$P_{\rm MC}^{\rm 1loop}$ in RPT, respectively. 
On the other hand, the third-order term $P_{ab}^{\rm(III)}$ is 
very similar to the term $P_{\rm MC}^{\rm 2loop}$ of 
\cite{CS2007}, but the factor $2$ is different. 
This discrepancy may be caused by the fact that  
the term $P^{\rm(III)}$ given above is associated with 
the higher-order contributions to the Born approximation of the 
mode-coupling function (\ref{eq:mode_coupling}), 
while the expression for $P_{\rm MC}^{\rm 2loop}$ has been 
properly derived as the leading-order contribution to 
the mode-coupling term at two-loop order. 
For the influence of the higher-order corrections,  
we leave the discussions in section \ref{subsec:results}.

\subsection{Approximate solution for non-linear propagator}
\label{subsec:approx_propagator}

Having obtained the analytic expressions for power spectra, 
our remaining task is to get the approximate solution for the 
non-linear propagator, $G_{ab}^{\rm approx}$, from the closure 
equation (\ref{eq:DIA_eq3}). Here, we just follow the procedure 
suggested by \citet{CS2006b} and construct the approximate 
solutions restricting their validity to the low-$k$ or the high-$k$ 
regions. Matching these asymptotic solutions appropriately 
at an intermediate regime, we obtain the global solutions for the 
non-linear propagator.

\subsubsection{Solutions at one-loop order}
\label{subsubsec:1-loop}

Let us first consider the low-$k$ limit of the non-linear propagator,  
where the perturbative treatment is safely applied. 
As it has been shown in section \ref{subsec:recovery}, 
our closure system consistently reproduces the one-loop results of 
the perturbation theory for power spectra $P_{ab}$. Indeed, this is 
also true for the propagator $G_{ab}$. 
Here, we explicitly write down the perturbative solutions 
at the one-loop level, $\delta G_{ab}^{\rm1\mbox{-}loop}(k|\eta,\eta')$. 
Using the linear propagator $g_{ab}$, 
perturbative solution of equation (\ref{eq:DIA_eq3}) is generally 
expressed as 
\begin{eqnarray}
\delta G^{\rm1\mbox{-}loop}_{ab}(k|\eta,\eta')
&=&4\int_{\eta'}^{\eta}d\eta_1 \,\,g_{ac}(\eta,\eta_1)\,
\int_{\eta'}^{\eta_1} d\eta_2
\int \frac{d^3\bfq}{(2\pi)^3}\,\,
\gamma_{cpq}(\bfq,\bfk-\bfq)\, 
\nonumber\\
&&\quad\quad\quad\quad\times
\gamma_{lrs}(-\bfq,\bfk)\,g_{ql}(\eta_1,\eta_2)\,
R_{pr}^{\rm lin}(q;\eta_1,\eta_2)\,g_{sb}(\eta_2,\eta'),
\label{eq:sol_1loop_propagator}
\end{eqnarray}
with the quantity $R_{ab}^{\rm lin}$ being the linear-order solution
for different-time power spectra.  In the approximation that 
the time-dependent matrix $\Omega_{ab}(\eta)$ given by equation 
(\ref{eq:matrix_M}) is replaced with the constant matrix 
with $\Omega_w=0$ and $f=1$, the analytical solution for linear 
propagator $g_{ab}$ satisfying the evolution equation 
(\ref{eq:eq_linear_propagator}) is obtained and 
is given by \citep[e.g.,][]{CS2006a,V2007a}: 
\begin{eqnarray}
g_{ab}(\eta_1,\eta_2)&=&
\Bigl\{
\frac{e^{\eta_1-\eta_2}}{5}\left(
\begin{array}{cc}
3 & 2 \\
3 & 2 
\end{array}
\right)+\frac{e^{-(3/2)(\eta_1-\eta_2)}}{5}\left(
\begin{array}{cc}
2 & -2 \\
-3 & 3 
\end{array}
\right)\Bigr\}\,\Theta(\eta_1-\eta_2) 
\label{eq:sol_linear_propagator}
\end{eqnarray}
with $\Theta(x)$ being the Heaviside step function. 
For the different-time spectrum $R_{ab}^{\rm lin}$, we have
\begin{eqnarray}
R_{ab}^{\rm lin}(k;\eta_1,\eta_2)&=&
e^{\eta_1+\eta_2}\left(
\begin{array}{cc}
1 & 1 \\
1 & 1 
\end{array}
\right)P_0(k), 
\label{eq:sol_linear_R_ab}
\end{eqnarray}
where we have neglected the contribution from the decaying mode.

Substituting the quantities (\ref{eq:sol_linear_propagator}) and 
(\ref{eq:sol_linear_R_ab}) into 
the next-to-leading order solution (\ref{eq:sol_1loop_propagator}), 
we first perform the time integrals over $\eta_1$ and $\eta_2$. 
As for the three-dimensional Fourier integral, we can write 
$d^3\bfq$ as $2\pi\, dq\,q^2\, dx$,  
where we have performed the integral over azimuthal angle. 
The variable $x$ is the cosine of the angle between $\bfk$ and $\bfq$, i.e., 
$x=\bfk\cdot\bfq/(kq)$, and the integral over $x$ is performed 
analytically. A straightforward but lengthy calculation leads to 
\citep[][]{CS2006b}: 
\begin{eqnarray}
\delta G_{ab}^{\rm1\mbox{-}loop}(k|\eta,\eta')=
\frac{e^{\eta-\eta'}}{5}\left(
\begin{array}{cc}
3\,X_{11} & 2\,X_{12} \\
3\,X_{21} & 2\,X_{22} 
\end{array}
\right)+\frac{e^{-(3/2)(\eta-\eta')}}{5}\left(
\begin{array}{cc}
2\,Y_{11} & -2\,Y_{12} \\
-3\,Y_{21} & 3\,Y_{22} 
\end{array}
\right),
\end{eqnarray}
where the matrices $X_{ab}$ and $Y_{ab}$ are given by 
\begin{eqnarray}
X_{ab}&=&e^{2\eta'}
\left(
\begin{array}{cc}
{\displaystyle \alpha(\eta-\eta')f(k)-\beta_g(\eta-\eta')i(k)}~ & ~
{\displaystyle \alpha(\eta-\eta')f(k)-\beta_g(\eta-\eta')h(k)} 
\\
{\displaystyle \alpha(\eta-\eta')g(k)+\gamma_g(\eta-\eta')h(k)}~ & ~
{\displaystyle \alpha(\eta-\eta')g(k)-\frac{3}{2}\,\gamma_g(\eta-\eta')i(k)} 
\end{array}
\right),
\label{eq:X_ab}
\\
 Y_{ab}&=&e^{2\eta'}
\left(
\begin{array}{cc}
{\displaystyle \delta(\eta-\eta')g(k)-\gamma_d(\eta-\eta')h(k)}~ & ~
{\displaystyle \delta(\eta-\eta')f(k)-\gamma_d(\eta-\eta')h(k)}
\\
{\displaystyle \delta(\eta-\eta')g(k)+\beta_d(\eta-\eta')i(k)}~ & ~
{\displaystyle \delta(\eta-\eta')f(k)-\frac{2}{3}\,\beta_d(\eta-\eta')h(k)}
\end{array}
\right).
\label{eq:Y_ab}
\end{eqnarray}
The above expressions contain  
time-dependent functions $\alpha$, $\beta_{g,d}$, $\gamma_{g,d}$ and 
$\delta$ and scale-dependent functions  $f$, $g$, $h$ and $i$, 
whose explicit expressions are summarized in Appendix 
\ref{appendix:func_propagator}. 
Note that the large-$k$ limit of the above functions satisfies 
\begin{equation}
f,~g,~h,~i\,\,\longrightarrow -\frac{1}{2} (k\,\sigma_{\rm v})^2
\label{eq:f_g_h_i}
\end{equation}
with the quantity $\sigma_{\rm v}$ being the velocity dispersion for 
linear fluctuation defined by 
\begin{equation}
\sigma_{\rm v}^2\equiv\frac{1}{3}\int\frac{d^3\bfq}{(2\pi)^3}\,
\frac{P_0(q)}{q^2}.
\label{eq:def_sigma_v}
\end{equation}

\subsubsection{Solutions in the high-$k$ limit}
\label{subsubsec:high-k}

Turn next to consider the high-$k$ limit of the non-linear propagator. 
In the evolution equation (\ref{eq:DIA_eq3}), 
we take the limit $\bfk\to+\infty$, while keeping $\bfq$ finite.   
In this limit, the vertex functions behave like
\begin{eqnarray}
\gamma_{cpq}(\bfq,\bfk-\bfq)\simeq \frac{1}{2}\,
\frac{\bfk\cdot\bfq}{|\bfq|^2}\,\delta_{cq}\,\delta_{2p},
\quad
\gamma_{lrs}(-\bfq,\bfk)\simeq -\,\frac{1}{2}\,
\frac{\bfk\cdot\bfq}{|\bfq|^2}\,\delta_{ls}\,\delta_{2r}.
\nonumber
\end{eqnarray}
To estimate the leading-order behavior analytically, 
the different-time spectrum 
$R_{pr}(q;\eta_1,\eta_2)$ in equation (\ref{eq:DIA_eq3}) 
is also treated approximately by replacing it with the linear-order 
quantity $R_{pr}^{\rm lin}$, given by equation(\ref{eq:sol_linear_R_ab}).

Then, the governing equation for non-linear propagator (\ref{eq:DIA_eq3}) 
is greatly simplified and we obtain
\begin{equation}
\widehat{\mbox{\boldmath$\Lambda$}}_{ab}(\eta)\,\,
G_{bc}(k|\eta,\,\eta')=-(k\sigma_{\rm v})^2\,
\int_{\eta'}^{\eta}d\eta''\,G_{ab}(k|\eta,\,\eta'')\,G_{bc}(k|\eta'',\,\eta')
\,e^{\eta+\eta''}, 
\label{eq:DIA_eq3_high-k}
\end{equation}
where the quantity $\sigma_{\rm v}$ is the rms fluctuation of the linear 
velocity given by equation (\ref{eq:def_sigma_v}). 
To solve the above equation, we adopt the following ansatz: 
\begin{eqnarray}
G_{ab}(k|\eta,\,\eta')=g_{ab}(\eta,\,\eta')\,f(k|\eta-\eta')
\nonumber
\end{eqnarray}
with the initial condition, $f(k|0)=1$. Using the basic property of 
the linear propagator, $g_{ab}(\eta,\eta')g_{bc}(\eta',\eta'')=
g_{ab}(\eta,\eta'')$, equation 
(\ref{eq:DIA_eq3_high-k}) is rewritten with 
\begin{eqnarray}
\frac{\partial}{\partial \tau}f(k|\tau)=-(k\sigma_{\rm v})^2
\int_{0}^{\tau}d\tau'\,f(k|\tau-\tau')\,f(k|\tau').  
\end{eqnarray}
Here, for convenience, we introduced the new time variable  
$\tau=e^{\eta}-e^{\eta'}$. The above equation has the analytical 
solution \citep[][]{V2007a}. 
Writing the Laplace transform of the function $f$ as 
$\widetilde{f}(k|s)=\int_0^{+\infty}ds\,e^{-s\tau}f(k|\tau)$, we have
\begin{eqnarray}
s\,\widetilde{f}(k|s)-1=-(k\sigma_{\rm v})^2\,[\widetilde{f}(k|s)]^2.  
\nonumber
\end{eqnarray}
The solution of this equation satisfying the limit,  
$\widetilde{f}\to0$ for $s\to+\infty$, becomes
\begin{eqnarray}
\widetilde{f}(k|s)=\frac{1}{2(k\sigma_{\rm v})^2}
\left[ -s+\sqrt{s^2+4(k\sigma_{\rm v})^2}\right].  
\nonumber
\end{eqnarray}
The inverse Laplace transform of 
the above expression is well-known and can be read off from 
the mathematical table: 
\begin{eqnarray}
f(k|\tau)= \frac{J_1(2x)}{x};\quad x=k\,\sigma_{\rm v}\,\tau
\end{eqnarray}
with the function $J_1(x)$ being a Bessel function of the first kind. 
Hence, the non-linear propagator in the high-$k$ limit finally becomes
\begin{equation}
G_{ab}(k|\eta,\,\eta')=g_{ab}(\eta,\,\eta')\,
\frac{J_1(2k\sigma_{\rm v}(e^{\eta}-e^{\eta'}))}
{k\sigma_{\rm v}(e^{\eta}-e^{\eta'})}.
\label{eq:G_high-k}
\end{equation}

\subsubsection{Matching the two solutions}
\label{subsubsec:matching}

\begin{figure}[t]
 \epsscale{1.12} \plottwo{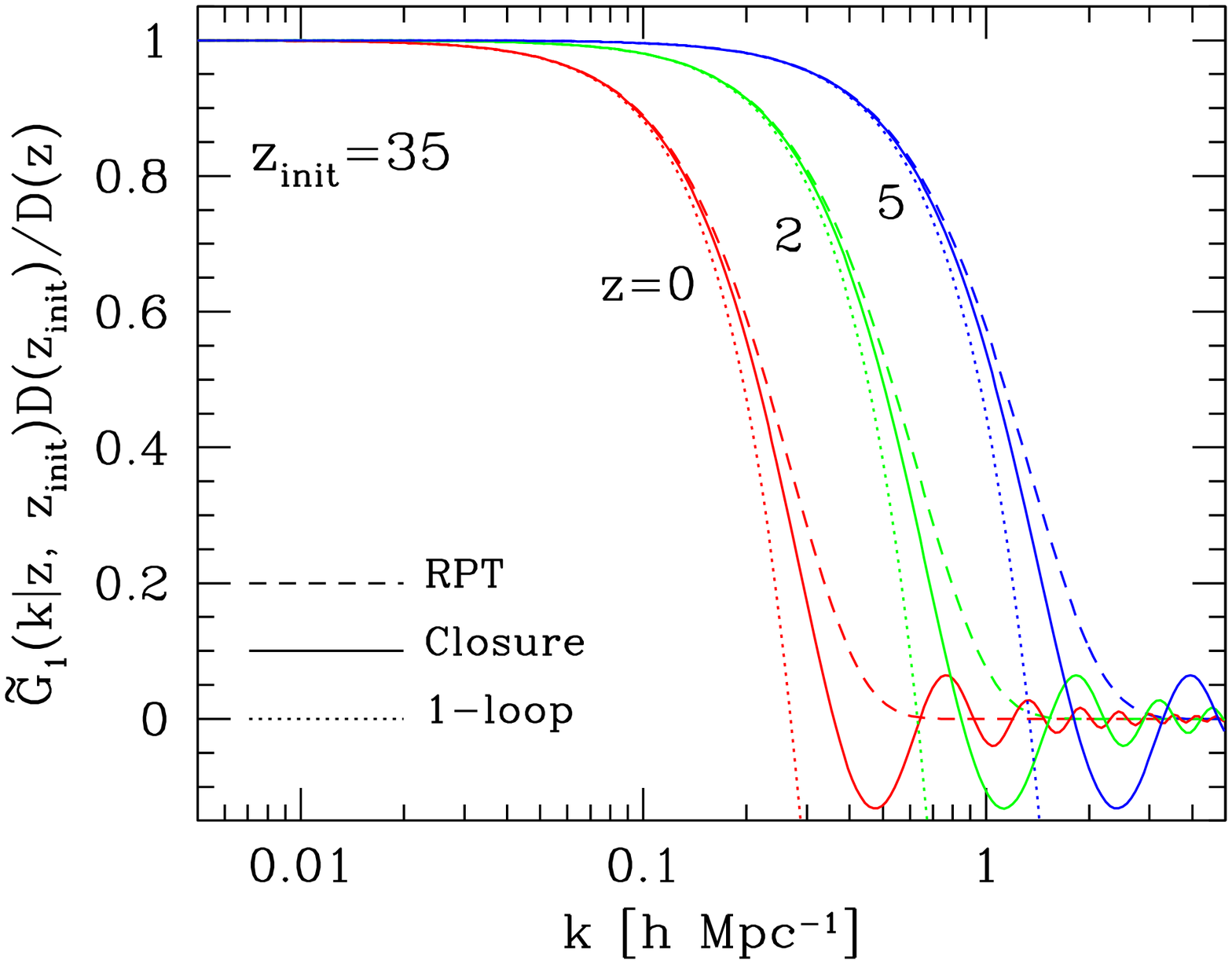}{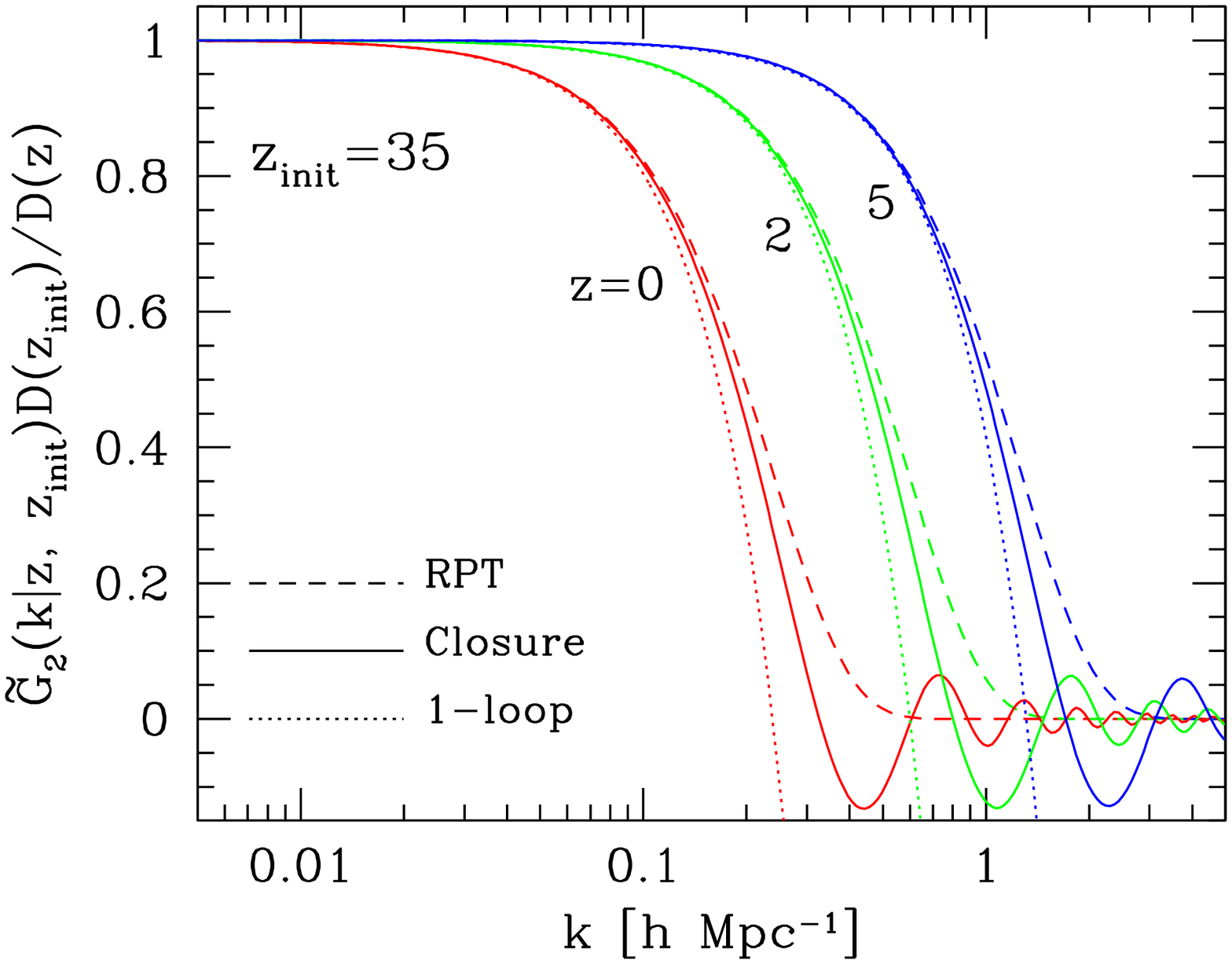} 
\caption{Approximate solutions for non-linear propagators 
  $\widetilde{G}_{1}(k|z,\,z_{\rm init})=G_{11}+G_{12}$ 
({\it left}) and $\widetilde{G}_{2}(k|z,\,z_{\rm init})=G_{21}+G_{22}$ 
({\it right}) as function of wave number. 
In each panel, 
solid lines show the results from closure theory, while the dashed 
lines are the propagators based on the renormalized perturbation theory (RPT) 
\citep{CS2006b}. For comparison, we also plot the results from 
the perturbation theory ($1$-loop). From left to right, the lines indicate 
$z=0,~2$ and $5$ with initial redshift $z_{\rm init}=35$.   
 \label{fig:G_delta_theta}}
\end{figure}

The asymptotic behaviors obtained in section \ref{subsubsec:1-loop} and 
\ref{subsubsec:high-k} have an overlapping region in which both of the 
approximations are applied. Therefore, matching these two solutions, one 
can obtain a global solution which 
would be a good approximation for the full propagator $G$. 
Let us recall from equation (\ref{eq:f_g_h_i}) 
that the propagator including the one-loop correction 
has the following asymptotic form: 
\begin{eqnarray}
g_{ab}(\eta,\eta')+\delta G_{ab}^{\rm1\mbox{-}loop}(k|\eta,\eta')
\,\,\stackrel{k\to\infty}{\longrightarrow}\,\,
g_{ab}(\eta,\eta')\,
\left\{1-
\frac{1}{2}\,(k\,\sigma_{\rm v})^2\,e^{2(\eta+\eta')}\right\}, 
\nonumber
\end{eqnarray}
where we have only considered the dominant terms at $\eta\to\infty$.  
On the other hand, the propagator in the high-$k$ limit, 
(\ref{eq:G_high-k}), is perturbatively expanded as 
\begin{eqnarray}
G_{ab}(k|\eta,\eta')\,\, \simeq \,\, g_{ab}(\eta,\eta') \, \left\{
1-\frac{x^2}{2}+\cdots
\right\}~;\quad x=k\,\sigma_{\rm v}\,(e^{\eta}-\eta^{\eta'}). 
\nonumber
\end{eqnarray}
Comparing these two expressions, 
the approximate solution smoothly matching these asymptotic 
behaviors at $\eta\to\infty$ may be 
\begin{eqnarray}
G_{ab}^{\rm approx}(k|\eta,\eta')=
\frac{e^{\eta-\eta'}}{5}\left(
\begin{array}{cc}
3\,P_{11} & 2\,P_{12} \\
3\,P_{21} & 2\,P_{22} 
\end{array}
\right)+\frac{e^{-(3/2)(\eta-\eta')}}{5}\left(
\begin{array}{cc}
2\,Q_{11} & -2\,Q_{12} \\
-3\,Q_{21} & 3\,Q_{22} 
\end{array}
\right). 
\label{eq:G_approx}
\end{eqnarray}
Here, the matrices $P_{ab}$ and $Q_{ab}$ are defined as 
\begin{equation}
P_{ab}=\frac{J_1(2\widetilde{X}_{ab})}{\widetilde{X}_{ab}},
\quad\quad
Q_{ab}=\frac{J_1(2\widetilde{Y}_{ab})}{\widetilde{Y}_{ab}}
\end{equation}
with $\widetilde{X}_{ab}\equiv|2X_{ab}|^{1/2}$ and 
$\widetilde{Y}_{ab}\equiv|2Y_{ab}|^{1/2}$ 
(see Eqs.[\ref{eq:X_ab}] and [\ref{eq:Y_ab}] 
for definitions of $X_{ab}$ and $Y_{ab}$).  
In the weakly non-linear regime, the propagator $G_{ab}^{\rm approx}$ 
correctly reproduces the one-loop results. In 
the large-$k$ limit, the function (\ref{eq:G_approx}) 
asymptotically approaches the solution (\ref{eq:G_high-k}).

Note that the approximate propagator (\ref{eq:G_approx}) is derived 
in the same way as done in RPT of \citet{CS2006b}, although 
the functional dependence is somewhat different because of the different 
high-$k$ behavior. In RPT, 
the matrices $P_{ab}$ and $Q_{ab}$ defined above should be replaced with 
\begin{eqnarray}
P_{ab}\to \exp(X_{ab}),\quad Q_{ab}\to \exp(Y_{ab}), 
\nonumber
\end{eqnarray}
which lead to the asymptotic behavior, 
$G_{ab}^{\rm(approx)}\to g_{ab}\exp(-x^2/2)$, in the high-$k$ limit.

Figure \ref{fig:G_delta_theta} shows the propagators 
$\widetilde{G}_1= G_{11}+G_{12}$ ({\it left}) and 
$\widetilde{G}_2= G_{21}+G_{22}$ ({\it right}) multiplying the 
factor $D(z_{\rm int})/D(z)$, for  
specific redshifts $z=0$, $2$ and $5$, 
and with initial redshift $z_{\rm init}=35$. 
The solid lines represent the approximate solutions 
obtained by matching the two asymptotic solutions.   
While the dotted lines show the results from the one-loop perturbation,  
the dashed lines indicate the non-linear propagators obtained from 
the RPT \citep[][]{CS2006b}.    
As increasing the wave number $k$, 
all the results exhibit the decaying behavior and the characteristic 
scale of the decay is shifted to low-$k$ as decreasing the redshift. 
A closer look at small scale (high-$k$) reveals that  
the one-loop propagators,  
$\widetilde{g}_{1,\,2}+\delta\widetilde{G}_{1,\,2}^{\rm1\mbox{-}loop}$,  
show unphysical behavior, which eventually become negative and 
tend to diverge. On the other hand,  the approximate solutions 
$\widetilde{G}_{1,\,2}^{\rm approx}$ obtained from 
the closure theory and RPT
asymptotically approach zero as $k\to\infty$. 
These damping behaviors are regarded as the 
non-perturbative effect as a result of the renormalization 
and/or self-consistent closure, which effectively 
takes account of the infinite series of higher-order corrections.  
Nevertheless, there exist some differences in the damping behaviors of the 
propagators.   
While the propagators in the RPT 
exhibit the exponential damping, 
the approximate solutions in closure theory show a damping oscillation.  
These differences may affect the final result of power spectrum. 
This point will be carefully discussed in next subsection.

\subsection{Results and discussion}
\label{subsec:results}

Having provided the basic ingredients for calculating the power 
spectrum, we now present the analytic results for power spectrum of 
density fluctuations, i.e., $P_{11}(k;z)$, 
and compare those with the results obtained from the 
RPT, particularly focusing on the characteristic scale of the BAOs.

\begin{figure}[t]
 \epsscale{0.7} \plotone{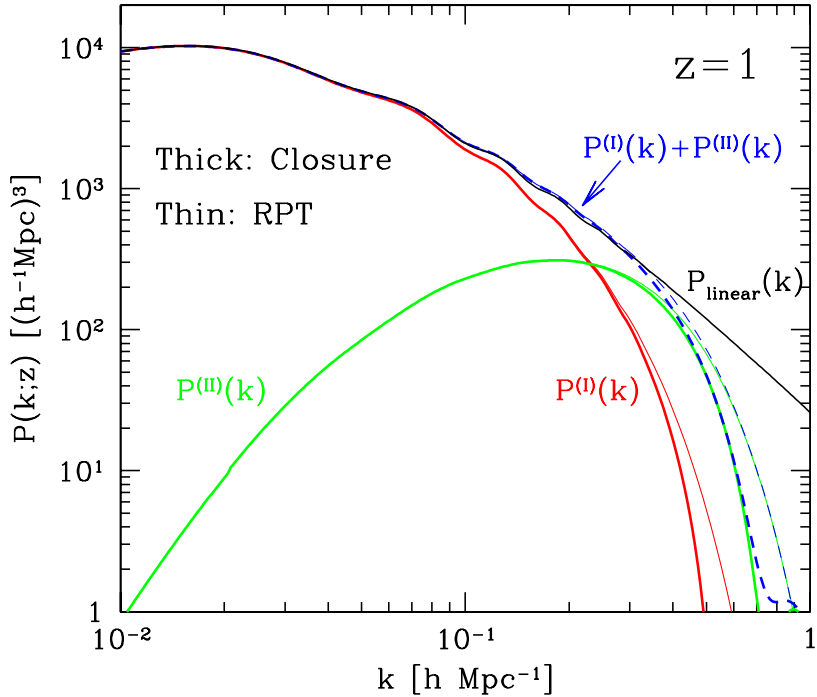}
\caption{Power spectrum of density fluctuations $P_{11}(k;z)$ at 
  $z=1$, obtained from the first-order Born approximation to the 
  integral solution (see Eq.[\ref{eq:Pk_Born}]). The contributions to 
  the total power spectrum are separately plotted as indicated 
  by $P^{\rm(I)}(k)$ and $P^{\rm(II)}(k)$ in the panel and 
  the total power spectrum, $P^{\rm(I)}(k)+P^{\rm(II)}(k)$, is depicted as 
  the dashed lines. Note 
  that in evaluating the power spectrum, the approximate solutions for 
  the non-linear propagators $G_{ab}^{\rm approx}$ were used. 
  Thick and thin lines indicate the results using the 
  approximate solutions $G_{ab}^{\rm approx}$ from the closure 
  theory and RPT, respectively. 
 \label{fig:Pk_BAO}}
\end{figure}
Figure \ref{fig:Pk_BAO} illustrates the overall behaviors of 
non-linear power spectrum of density fluctuations 
$P(k;z)\equiv P_{11}(k;z)$ given at $z=1$, 
based on the Born approximation (\ref{eq:Pk_Born}).  Here, 
the contributions to the total power spectrum up to the first-order 
Born approximation, i.e., $P^{\rm(I)}(k)$ and $P^{\rm(II)}(k)$, are 
separately plotted. Thin and thick lines 
represent the results from RPT and the 
closure theory, respectively.  The result from RPT is basically the same one 
as presented by \citet{CS2007}, although they further considered 
the higher-order contribution coming from the two-loop correction.  
Due to the damping behavior in the 
non-linear propagators, each contribution to the total power spectrum 
rapidly falls off in both predictions 
and their amplitudes become significantly lower than the linear power 
spectrum on small scales (labeled by $P_{\rm linear}(k)$), 
where the differences between the two predictions become manifest. 
Turning to focus on the scales larger than the damping scale, 
the contribution coming from $P^{\rm(II)}$ 
becomes maximum around $k\sim0.2h\,\mbox{Mpc}^{-1}$, where the non-linear 
enhancement of the power spectrum $P(k)$ can be seen and the differences 
between closure theory and RPT become 
fairly small. In particular, for the scale of our interest on 
BAOs around $k\simlt0.01-0.3h\,\mbox{Mpc}^{-1}$, 
one cannot clearly distinguish between both predictions from 
Figure \ref{fig:Pk_BAO}.

\begin{figure}[t]
 \epsscale{1.1} \plottwo{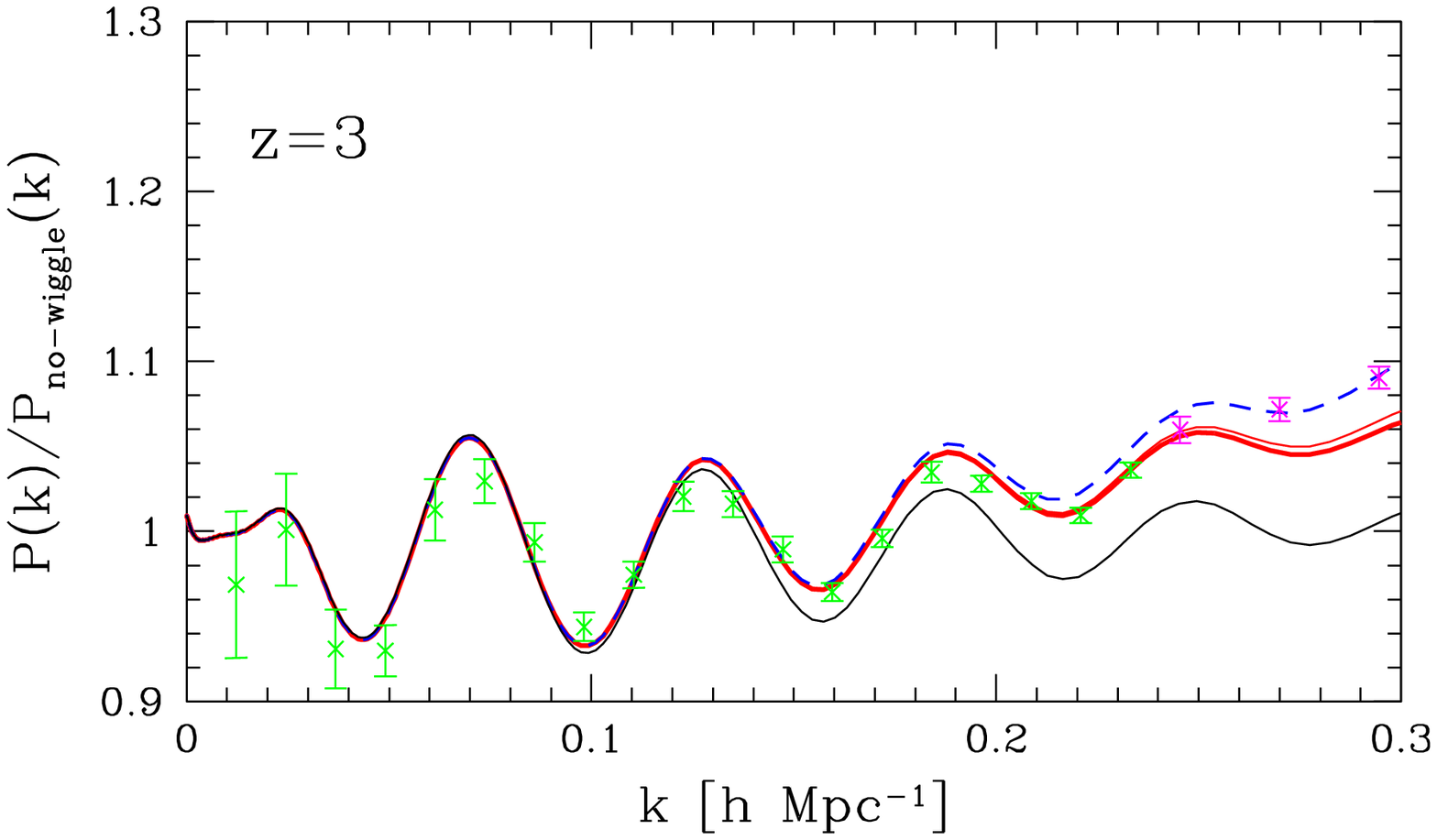}{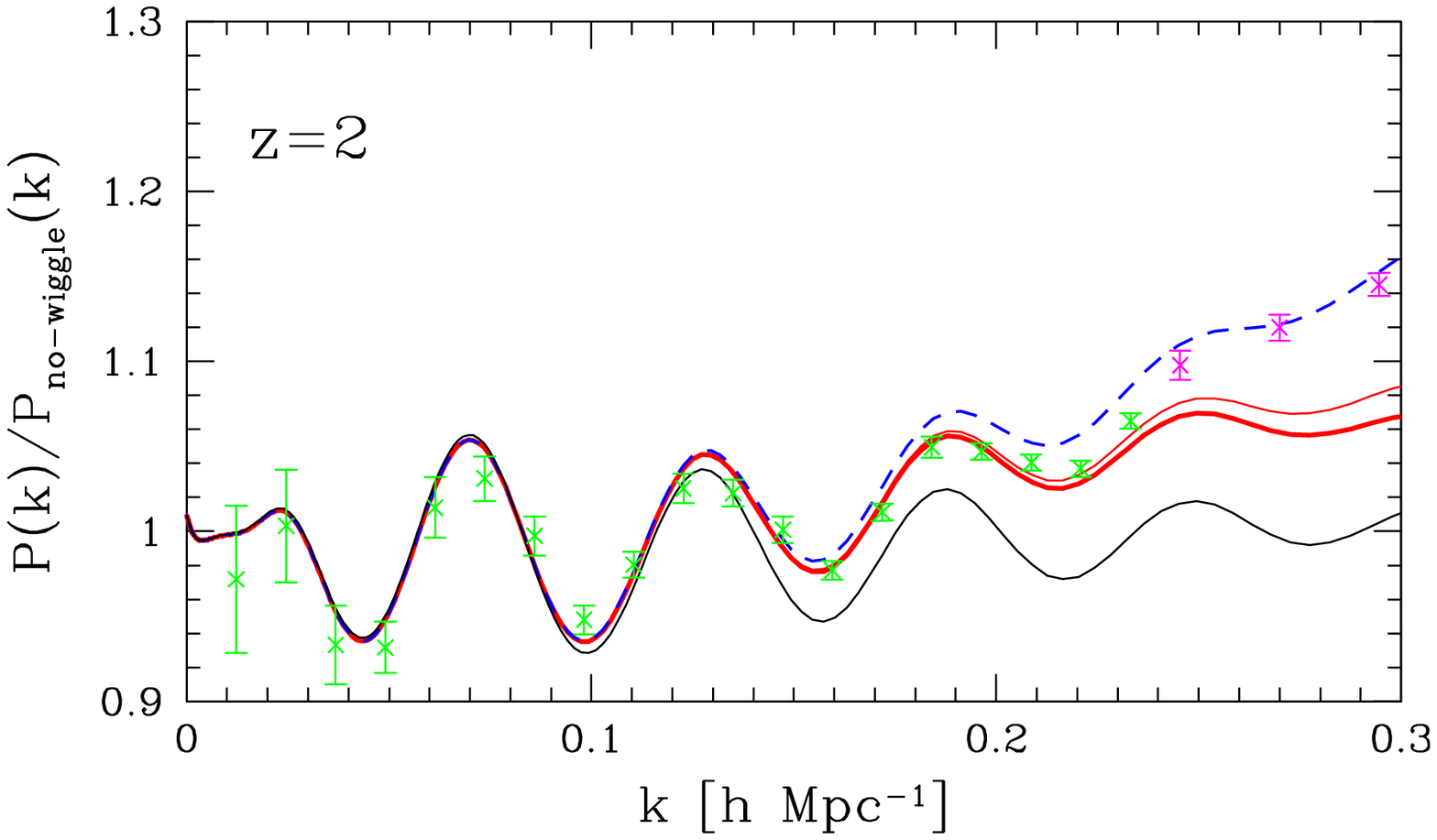} 

 \vspace*{0.5cm}

 \epsscale{1.1} \plottwo{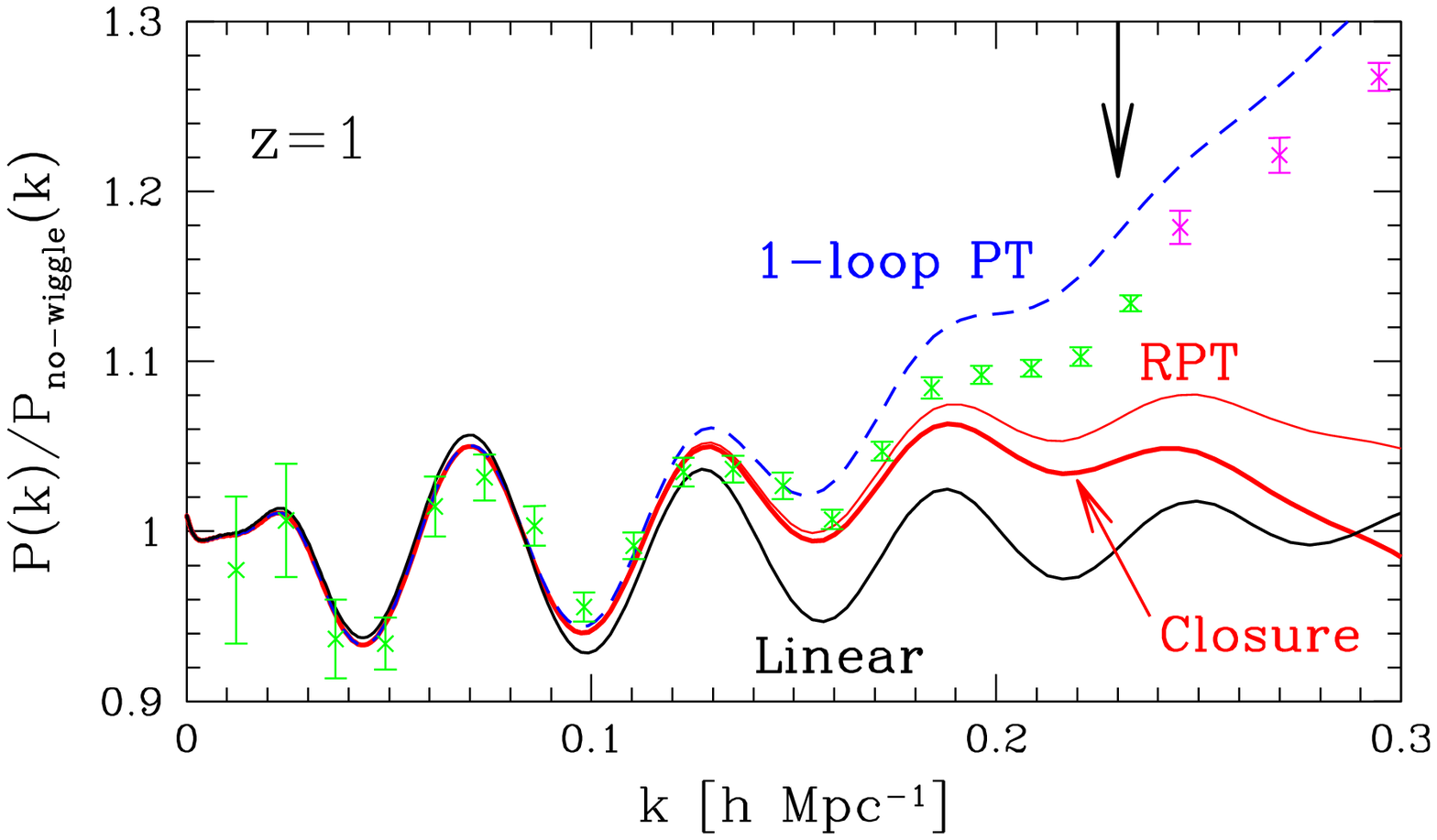}{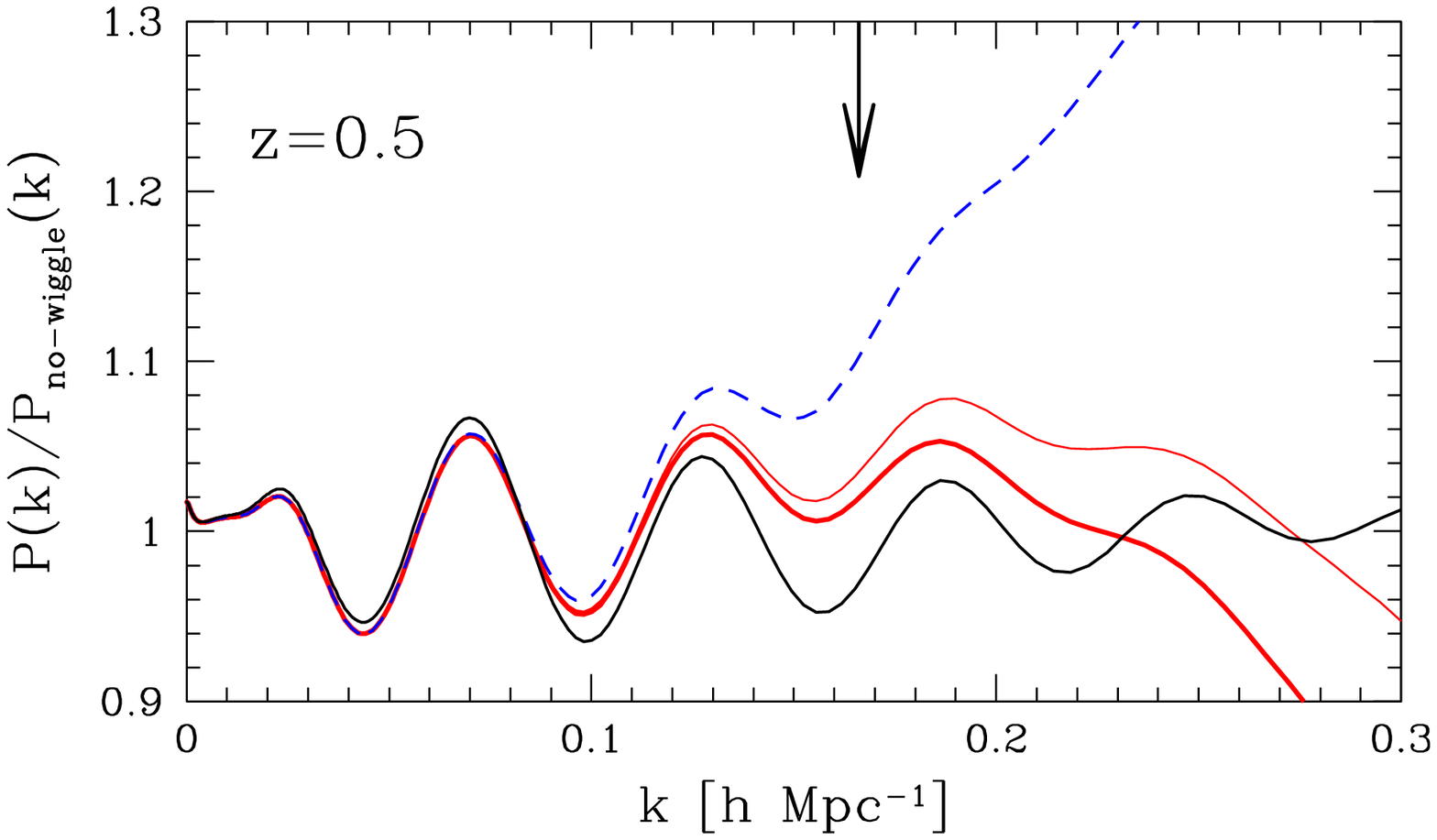} 
\caption{Ratio of non-linear power spectrum to smoothed linear 
  spectrum, $P(k)/P_{\rm no\mbox{-}wiggle}(k)$, 
  given at specific redshifts, $z=3$, $2$, $1$ and $0.5$. 
  The error bar represents the N-body results taken from \citet{JK2006},  
  in which different color indicates the results with different 
  box size (see their paper in detail). 
  Here, smoothed linear spectra $P_{\rm no\mbox{-}wiggle}(k)$ 
  were calculated from the 
  linear transfer function without baryon acoustic oscillation 
  according to the fitting formula of \citet{EH1998} 
  (Eq.[29] of their paper). The non-linear power spectra 
  are obtained from the first-order Born approximation to the integral 
  solution (Eq.[\ref{eq:Pk_Born}]), with approximate solutions of the 
  non-linear propagator given by closure theory (thick) and RPT (thin). 
  For comparison, one-loop predictions from the standard perturbation 
  theory are plotted in dashed lines. Also, in panels with $z=1$ and $0.5$, 
  maximum wave number for limitation of one-loop perturbation is indicated 
  by vertical arrows, according to the criterion, 
  $\Delta^2(k)\equiv k^3P(k)/(2\pi^2)\simlt0.4$ \citep{JK2006}. 
 \label{fig:ratio_Pk}}
\end{figure}
\begin{figure}[t]
 \epsscale{1.1} \plottwo{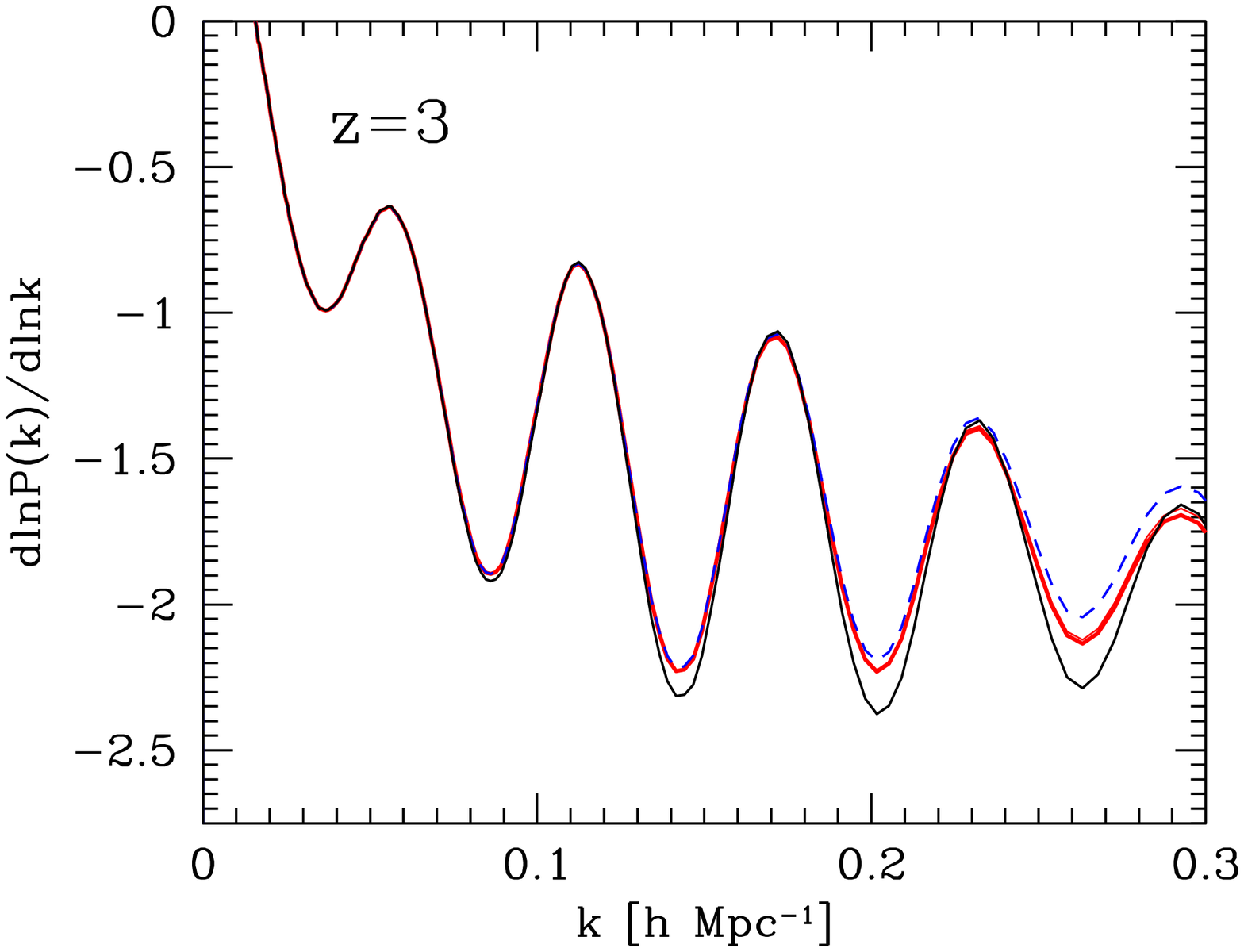}{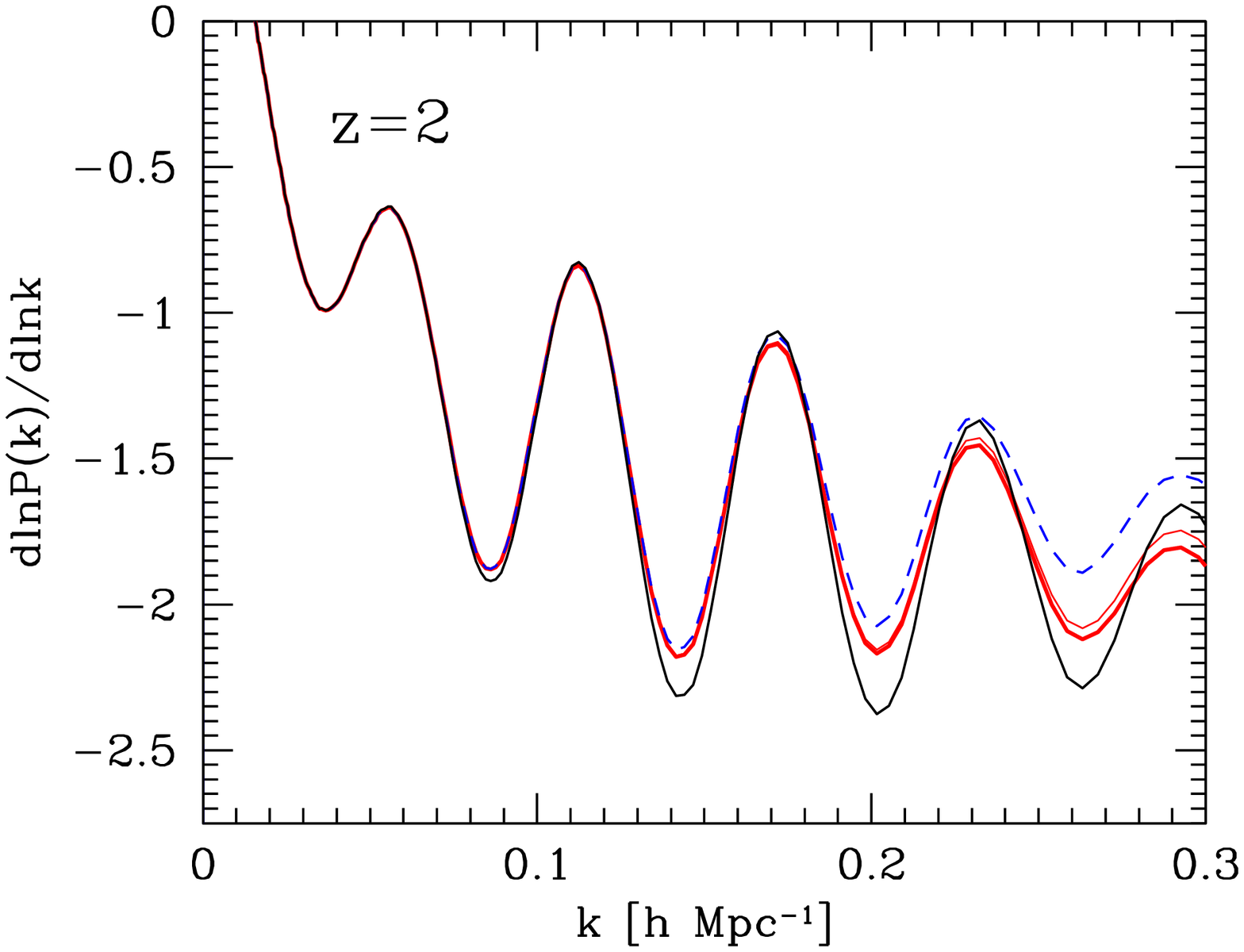} 

 \vspace*{0.5cm}

 \epsscale{1.1} \plottwo{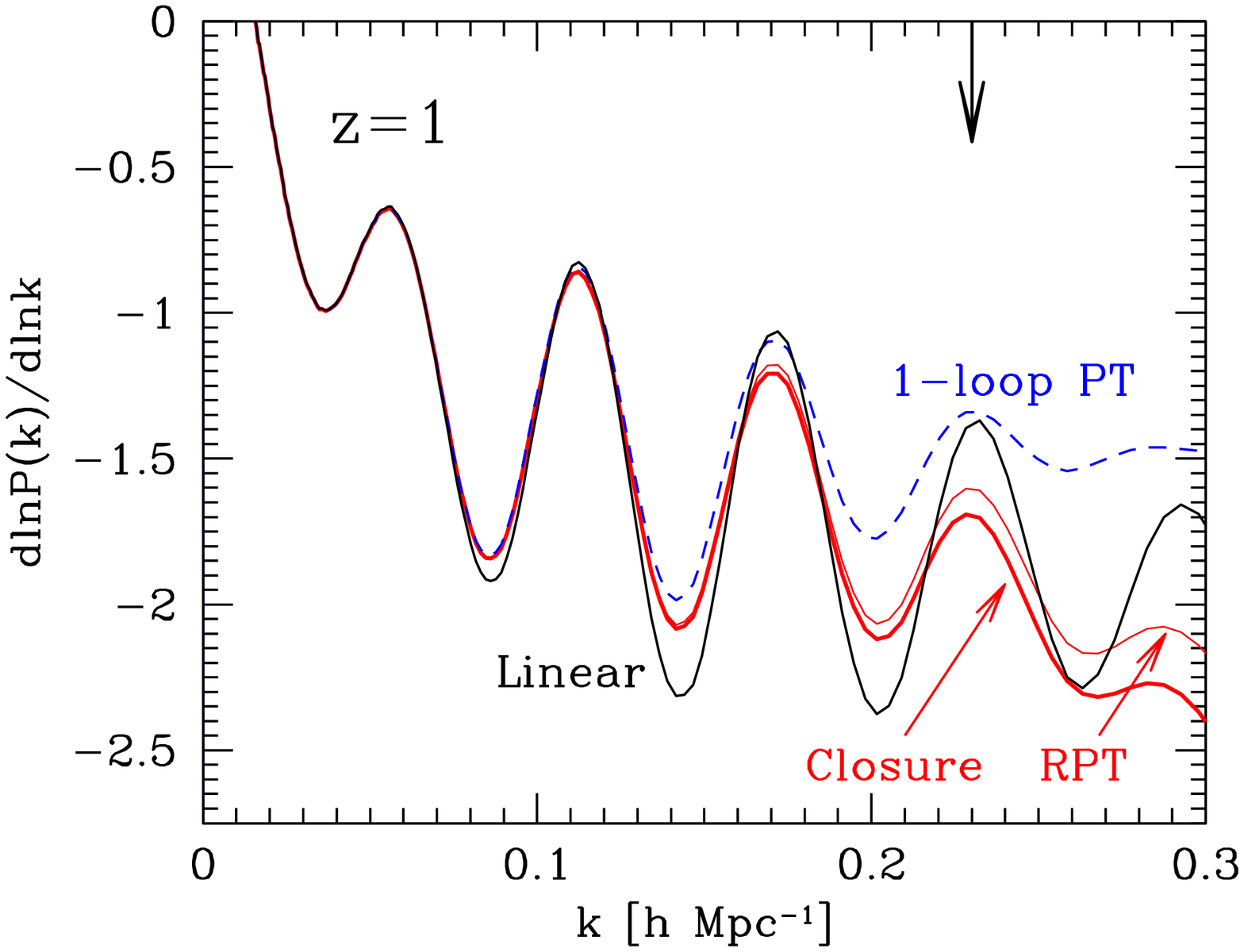}{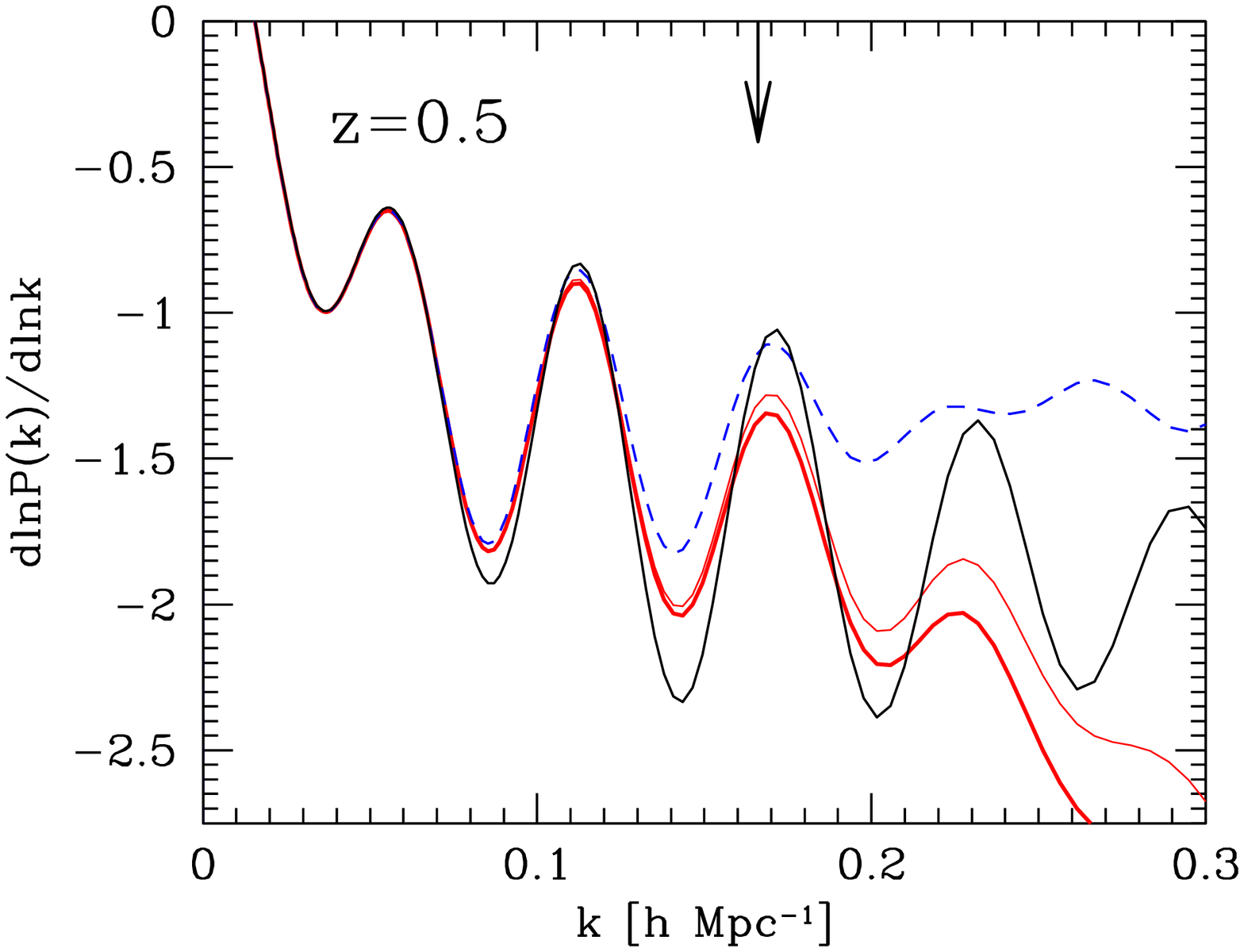} 
\caption{Logarithmic derivative of non-linear power spectrum, 
  $d\ln P(k)/d\ln k$, at redshifts $z=3$, $2$, $1$ and $0.5$.  
  Line types and labels are the same as in Fig.\ref{fig:ratio_Pk}. 
 \label{fig:dlnPdlnk}}
\end{figure}

To enlarge the differences between the two predictions and to clarify 
the non-linear behaviors of the BAOs, 
in Figure \ref{fig:ratio_Pk}, snapshots of the power spectra 
divided by the smooth linear spectrum, 
$P(k)/P_{\rm no\mbox{-}wiggle}(k)$ are plotted for 
specific redshifts $z=3$, $2$, $1$ and $0.5$ together with the 
$N$-body results kindly provided by \citet{JK2006} (except for z=0.5), 
while in Figure 
\ref{fig:dlnPdlnk}, we present the logarithmic derivative of the 
power spectra, $d\ln P(k)/d\ln k$. All the results are plotted 
in linear scales. The smooth 
power spectra, $P_{\rm no\mbox{-}wiggle}(k)$, 
was calculated from the linear transfer function without 
BAOs based on the fitting formula of \citet{EH1998} 
(see Eq.[29] in their paper). Note that 
the power spectra calculated from the closure theory and RPT 
are the sum of the leading-order 
contributions, $P^{\rm(I)}(k)$ and $P^{\rm(II)}(k)$, not including 
the higher-order term $P^{\rm(III)}(k)$. 
For comparison, the one-loop predictions from the standard perturbation 
theory are also depicted as dashed lines.

On large scales (low-$k$), the predictions both from 
the closure theory and RPT reasonably match the one-loop results of standard 
perturbation theory, as anticipated. 
This is just the quantitative check for the 
non-perturbative methods discussed in section \ref{subsec:recovery}. 
On the other hand, on smaller scales (high-$k$), 
the deviations from the one-loop 
perturbation become manifest and the amplitude of the 
predictions both from closure theory and RPT 
is suppressed compared to the one-loop predictions. 
The reduction of  the power spectrum amplitude is the 
natural outcome of the damping behaviors 
appearing in the non-linear propagator and 
it qualitatively explains the behaviors seen in the N-body 
simulations \citep[e.g.,][]{JK2006,MP2007,CS2007}. However, 
at lower redshifts $z=1$ and $z=0.5$, the suppression of the amplitude 
is so significant that the predictions eventually become lower than 
the linear theory prediction. Compared to the prediction from 
RPT, the suppression of the amplitude 
is even larger for the prediction from the closure theory.

The strong suppression seen in the low redshifts seems somewhat unphysical,  
indicating the failure of our present analysis. 
In the panels with redshift $z=1$ and $z=0.5$ of 
Figures \ref{fig:ratio_Pk} and \ref{fig:dlnPdlnk}, 
the maximum wave number for the limitation of one-loop perturbation 
theory is estimated according to the condition,  
$\Delta^2(k)\equiv k^3P(k)/(2\pi^2)\simlt0.4$, given by 
\citet{JK2006} and is depicted as vertical arrows, 
above which the perturbation prediction fails to reproduce the N-body 
results within the $1\%$ accuracy. Comparaed to those 
critical scales, the validity range of the 
non-perturbative predictions seems rather narrow. 
While this statement seems rather controversial, our present analysis   
actually relies on several approximate treatments. 
Hence, the inclusion of the 
higher-order terms such as $P^{\rm(III)}(k)$ might be essential for 
the correct non-linear behaviors. 
\citet{CS2007} reported that the contributions higher than the 
two-loop correction become important on small scales 
and the prediction including 
the two-loop correction does improve the prediction, which 
reasonably agrees with N-body results within an accuracy of the  
N-body calculation. As we noted previously, however, 
the two-loop correction evaluated by the first-order Born approximation is 
comparable to the higher-order Born approximation coming from 
the $P^{\rm(III)}(k)$ term, and at a level of the present analysis, 
it is not clear 
whether the higher-loop corrections rather than the higher-order Born 
approximation of the one-loop correction are essential or not. 
Rather, one obvious thing is that 
a further improvement of the approximate treatment is necessary   
to faithfully predict the non-linear evolution of BAOs. 
In the line of this, 
self-consistent treatment to solve the evolution equations 
(\ref{eq:DIA_eq1}), (\ref{eq:DIA_eq2}) and (\ref{eq:DIA_eq3}),  
would be one plausible approach, which we will address this issue  
in next task.

\section{Conclusions}
\label{sec:conclusion}

In this paper, motivated by a forthcoming experiment on 
the precision measurement of the 
BAOs imprinted on the matter power spectrum,  
a new theoretical tool for predicting the non-linear gravitational 
evolution of power spectrum was presented. In particular, 
we have applied the non-linear statistical method in turbulence to 
the cosmological perturbation theory 
and derived a closed set of matter power spectrum.  
The resultant evolution equations 
(\ref{eq:DIA_eq1})--(\ref{eq:DIA_eq3}) with Fourier kernels 
(\ref{eq:kernel_F2}) and (\ref{eq:kernel_K2}) are the non-linear 
coupled system of integro-differential equations and these   
consistently recover the 
one-loop results of standard perturbation theory. 
Further, the exact integral expressions for the solutions of 
closure equations were 
obtained (see Eqs.[\ref{eq:integral_1}] and [\ref{eq:integral_2}] with  
mode-coupling function [\ref{eq:mode_coupling}]), 
whose analytical expressions coincide with the renormalized 
one-loop results of the theory developed by \citet{CS2006a}, apart from 
the corrections coming from the vertex renormalization.

Based on the Born approximation to the exact integral expressions, 
we next tried to evaluate the non-linear power spectrum analytically. 
Constructing an approximate solution for non-linear 
propagator, which smoothly matches the two asymptotic solutions valid at 
low-$k$ and high-$k$, power spectrum of density fluctuations was 
computed and the results are compared with those obtained from 
the RPT. Due to the non-linear damping 
behavior of the non-linear propagator, the resultant amplitude of the 
power spectra up to the first-order Born approximation is strongly 
suppressed on small scales and this effect becomes significant as 
decreasing redshift. As a result, predicted spectra at high-$k$ 
both from the closure theory and RPT fail to reproduce the 
N-body trends and the inclusion of the higher-order corrections 
is required. 
Nevertheless, 
at the intermediate scales, where the damping behavior of the non-linear 
propagator is rather mild, the closure theory and RPT both predict the 
deviation from the 
one-loop results of standard perturbation theory, which qualitatively 
agree with the N-body results.

Although the analytical treatment 
presented here is still primitive and the range of the applicability is 
severely restricted by the validity of the approximations, 
the results indicate that our closure theory is a promising 
non-perturbative approach 
comparable to the RPT and a more elaborated study 
will provide an accurate prediction for non-linear power spectra 
going beyond standard perturbation theory. In this respect, 
a direct numerical treatment of the closure equations 
(\ref{eq:DIA_eq1})--(\ref{eq:DIA_eq3}) is our urgent task. 
As it has been reported by \citet{V2007a}, 
thanks to the full numerical treatment, the evolved result of the 
power spectrum shows several desirable properties and it 
qualitatively matches the N-body trends even at high-$k$. In future 
publication, 
these points will be further investigated from a quantitative point-of-view, 
particularly focusing on the non-linear evolution of BAOs.

Finally, one criticism on the present approach including the 
currently existing non-perturbative methods  
is that the present methodology heavily relies  
on the single-stream approximation of the Vlasov equation. In a 
strongly non-linear regime at  
the high-$k$ region, the shell-crossing eventually occurs and the fluid 
description would be broken. A preliminary investigation 
suggests that the break down of the single-stream approximation 
may arise at $k\sim1-2\,h\mbox{Mpc}^{-1}$ \citep[][]{S2000,A2007}. 
Therefore, for the high-$k$ region beyond the critical scale, the present 
approach cannot be applied and a more delicate 
treatment based on the Vlasov equation must be developed. This issue 
would be particularly crucial for accurate theoretical predictions to  
the cosmic shear statistics.

\bigskip
We are grateful to Keisuke Izumi, Shun Saito, Misao Sasaki, Jiro Soda and 
Yasushi Suto for useful discussions.  We also thank Eiichiro Komatsu for 
providing the N-body data of \citet{JK2006}. 
This work is supported in part by a Grant-in-Aid for Scientific Research 
from the Japan Society for Promotion of Science (No.~18740132).

\appendix
\section{One-loop perturbation}
\label{appendix:one-loop_PT}

Here, we show that linear plus one-loop power spectra, 
$P_{ab}^{(11)}(k)+P_{ab}^{(22)}(k)+P_{ab}^{(13)}(k)$, 
obtained from the standard perturbation theory indeed satisfy the 
evolution equations given by (\ref{eq:recovery_1loopPT}).
To do this, we first write down the evolution equations for perturbed 
quantities $\Phi_{a}^{(i)}(k;\eta)$ in each order:  
\begin{eqnarray}
\widehat{\mbox{\boldmath$\Lambda$}}_{ab}(\eta)\Phi_b^{(1)}(\bfk;\eta) &=& 0,
\label{eq:LPhi1} \\
\widehat{\mbox{\boldmath$\Lambda$}}_{ab}(\eta)\Phi_b^{(2)}(\bfk;\eta) &=& 
\int\frac{d^3\bfk_1d^3\bfk_2}{(2\pi)^3}\,
\delta_D(\bfk-\bfk_1-\bfk_2)
\,\gamma_{apq}(\bfk_1,\bfk_2)\,
\Phi_p^{(1)}(\bfk_1;\eta) \Phi_q^{(1)}(\bfk_2;\eta),
\label{eq:LPhi2} \\
\widehat{\mbox{\boldmath$\Lambda$}}_{ab}(\eta)\Phi_b^{(3)}(\bfk;\eta) &=& 2
\int\frac{d^3\bfk_1d^3\bfk_2}{(2\pi)^3}\,
\delta_D(\bfk-\bfk_1-\bfk_2)
\,\gamma_{apq}(\bfk_1,\bfk_2)\,
\Phi_p^{(1)}(\bfk_1;\eta) \Phi_q^{(2)}(\bfk_2;\eta).
\label{eq:LPhi3}
\end{eqnarray}
The above equations are formally solved with a help of 
the linear propagator, $g_{ab}(\eta-\eta')$. For instance, 
the solution for second-order quantity $\Phi_a^{(2)}$ can be written as 
\begin{eqnarray}
\Phi_a^{(2)}(\bfk;\eta)&=&\int_{\eta_0}^{\eta}d\eta' \,\,
g_{ab}(\eta-\eta')\,
\nonumber\\
&&\times
\int\frac{d^3\bfk_1d^3\bfk_2}{(2\pi)^3}\,\,
\delta_D(\bfk-\bfk_1-\bfk_2)\,\,\gamma_{bpq}(\bfk_1,\bfk_2)\,
\Phi_p^{(1)}(\bfk_1;\eta') \Phi_q^{(1)}(\bfk_2;\eta').
\label{eq:sol_Phi2}
\end{eqnarray}

Now, let us consider the time evolution of power spectrum, 
$P^{(22)}(k;\eta)$. Acting the operator 
$\widehat{\mathbf{\Sigma}}_{abcd}(\eta)$ 
defined by (\ref{eq:def_of_Sigma_Lambda}) on the ensemble average 
$\langle\Phi_c^{(2)}(\bfk;\eta)\Phi_d^{(2)}(\bfk';\eta)\rangle$,  
we obtain 
\begin{eqnarray}
&&\widehat{\mathbf{\Sigma}}_{abcd}(\eta)\,\Bigl\langle 
\Phi_c^{(2)}(\bfk;\eta)\Phi_d^{(2)}(\bfk';\eta)\Bigr\rangle
\nonumber\\
&&\quad\quad= 
\Bigl\langle
\Phi_a^{(2)}(\bfk;\eta)\,\,
\left\{\widehat{\mathbf{\Lambda}}_{bd}(\eta)\,\Phi_d^{(2)}(\bfk';\eta)\right\}
\Bigr\rangle
+
\Bigl\langle 
\left\{\widehat{\mathbf{\Lambda}}_{ac}(\eta)\,
\Phi_c^{(2)}(\bfk;\eta)\right\}\,\,\Phi_b^{(2)}(\bfk';\eta)
\Bigr\rangle
\nonumber\\
&& \quad\quad=\int_{\eta_0}^{\eta}d\eta' \,g_{al}(\eta-\eta')
\int\frac{d^3\bfk_1d^3\bfk_2}{(2\pi)^3}
\int\frac{d^3\bfk_3d^3\bfk_4}{(2\pi)^3}\,\,
\delta_D(\bfk-\bfk_1-\bfk_2)\,\delta_D(\bfk'-\bfk_3-\bfk_4)\,
\nonumber\\
&&\quad\quad\quad\quad\times\,\gamma_{lrs}(\bfk_1,\bfk_2)\,
\gamma_{bpq}(\bfk_3,\bfk_4)\,
\Bigl\langle \Phi_r^{(1)}(\bfk_1;\eta')\Phi_s^{(1)}(\bfk_2;\eta')
\Phi_p^{(1)}(\bfk_3;\eta)\Phi_q^{(1)}(\bfk_4;\eta)\Bigr\rangle
\nonumber\\
&&\quad\quad\quad+ 
\quad\Bigl(\quad a\longleftrightarrow b, \quad \quad 
\bfk \longleftrightarrow \bfk' \quad\Bigr),
\nonumber\\
&& \quad\quad=2\int_{\eta_0}^{\eta}d\eta' \, g_{al}(\eta-\eta')
\int d^3\bfk_1d^3\bfk_2\,\,
\delta_D(\bfk'-\bfk_1-\bfk_2)\,\delta_D(\bfk+\bfk_1+\bfk_2)\,
\nonumber\\
&&\quad\quad\quad\quad\times\,
\gamma_{bpq}(\bfk_1,\bfk_2)\gamma_{lrs}(\bfk_1,\bfk_2)
R_{pr}^{(11)}(k_1;\eta,\eta')R_{qs}^{(11)}(k_2;\eta,\eta'),
\nonumber\\
&&\quad\quad\quad+ 
\quad\Bigl(\quad a\longleftrightarrow b, \quad \quad 
\bfk \longleftrightarrow \bfk' \quad\Bigr),
\label{eq:Sigma22}
\end{eqnarray}
where, in the last equality, we have replaced the four-point functions 
with a product of the two-point functions $R_{ab}^{(11)}$ 
according to Wick's theorem. The quantity $R_{ab}^{(11)}$ is the 
linear cross spectrum defined by 
\begin{eqnarray}
\Bigl\langle\Phi_a^{(1)}(\bfk;\eta)\Phi_b^{(1)}(\bfk';\eta')\Bigr\rangle 
= (2\pi)^3\,\delta_D(\bfk+\bfk')\,\,R_{ab}^{(11)}(k;\eta,\eta'),\quad
(\eta\geq \eta').
\label{eq:def_R11}
\end{eqnarray}
Integrating equation (\ref{eq:Sigma22}) over the Fourier mode $\bfk'$ 
leads to 
\begin{eqnarray}
\widehat{\mathbf{\Sigma}}_{abcd}(\eta)\,P_{cd}^{(22)}(k;\eta)
&=&
2\int_{\eta_0}^{\eta}d\eta' \,g_{al}(\eta-\eta')
\int\frac{d^3\bfk_1d^3\bfk_2}{(2\pi)^3}\,\,
\delta_D(\bfk+\bfk_1+\bfk_2)\,\,
\nonumber\\
&&\quad\quad\quad\times\,\gamma_{lrs}(\bfk_1,\bfk_2)\,
\gamma_{bpq}(\bfk_1,\bfk_2)\,
R_{pr}^{(11)}(k_1;\eta,\eta')R_{qs}^{(11)}(k_2;\eta,\eta')
\nonumber\\
&+&\quad\Bigl(\quad a\longleftrightarrow b, \quad \quad 
\bfk \longleftrightarrow -\bfk \quad\Bigr).
\label{eq:evolve_P22}
\end{eqnarray}

Next consider the evolution equation for power spectrum $P^{(13)}(k;\eta)$. 
Repeating the similar calculations given above, we have 
\begin{eqnarray}
&&\widehat{\mathbf{\Sigma}}_{abcd}(\eta)\,\Bigl\langle 
\Phi_c^{(1)}(\bfk;\eta)\Phi_d^{(3)}(\bfk';\eta)+
\Phi_c^{(3)}(\bfk;\eta)\Phi_d^{(1)}(\bfk';\eta) \Bigr\rangle
\nonumber\\
&&\quad\quad= \Bigl\langle \Phi_a^{(1)}(\bfk;\eta)\,\,
\left\{\widehat{\mathbf{\Lambda}}_{bd}(\eta)\,\Phi_d^{(3)}(\bfk';\eta)\right\}
\Bigr\rangle+
\Bigl\langle 
\left\{\widehat{\mathbf{\Lambda}}_{ac}(\eta)\,\Phi_c^{(3)}(\bfk;\eta)\right\}
\,\,\Phi_b^{(1)}(\bfk';\eta)
\Bigr\rangle
\nonumber\\
&&\quad\quad
=2\int_{\eta_0}^{\eta}d\eta'
\int\frac{d^3\bfk_1d^3\bfk_2}{(2\pi)^3}
\int\frac{d^3\bfk_3d^3\bfk_4}{(2\pi)^3}\,\,
\delta_D(\bfk'-\bfk_1-\bfk_2)\,\delta_D(\bfk_2-\bfk_3-\bfk_4)
\nonumber\\
&&\quad\quad\quad\quad\times\,
g_{ql}(\eta-\eta')\,\gamma_{bpq}(\bfk_1,\bfk_2)\,\gamma_{lrs}(\bfk_3,\bfk_4)
\Bigl\langle\Phi_a^{(1)}(\bfk;\eta)
\Phi_p^{(1)}(\bfk_1;\eta)\Phi_r^{(1)}(\bfk_3;\eta')\Phi_s^{(1)}(\bfk_4;\eta')
\Bigr\rangle
\nonumber\\
&&\quad\quad\quad\quad+ 
\quad\Bigl(\quad a\longleftrightarrow b, \quad \quad 
\bfk \longleftrightarrow \bfk' \quad\Bigr)
\nonumber \\
&& \quad\quad=4\int_{\eta_0}^{\eta}d\eta' \, g_{ql}(\eta-\eta')
\int d^3\bfk_1d^3\bfk_2\,\,
\delta_D(\bfk'-\bfk_1-\bfk_2)\,\delta_D(\bfk+\bfk_1+\bfk_2)\,
\nonumber\\
&&\quad\quad\quad\quad\times\,
\gamma_{bpq}(\bfk_1,\bfk_2)\gamma_{lrs}(\bfk,\bfk_1)
R_{ar}^{(11)}(k;\eta,\eta')R_{ps}^{(11)}(k_1;\eta,\eta')
\nonumber\\
&&\quad\quad\quad+ 
\quad\Bigl(\quad a\longleftrightarrow b, \quad \quad 
\bfk \longleftrightarrow \bfk' \quad\Bigr).
\nonumber
\end{eqnarray}
Here, in the second equality of the above equation, 
we have used equations (\ref{eq:LPhi1}) and 
(\ref{eq:LPhi3}), and substituted the formal solution 
(\ref{eq:sol_Phi2}) to rewrite all the perturbed 
quantities with the linear-order one, $\Phi_b^{(1)}(\mathbf{k},\eta)$. 
Then, integrating over $\bfk'$, we obtain
\begin{eqnarray}
\widehat{\mathbf{\Sigma}}_{abcd}P_{cd}^{(13)}(k;\eta) &=&
4\int_{\eta_0}^{\eta}d\eta' \, g_{ql}(\eta-\eta')\,
\int\frac{d^3\bfk_1d^3\bfk_2}{(2\pi)^3}\,\,
\delta_D(\bfk+\bfk_1+\bfk_2)\,\,
\nonumber\\
&&\quad\quad\quad\times\,\,\gamma_{bpq}(\bfk_1,\bfk_2)\,
\gamma_{lrs}(\bfk,\bfk_1)\,
R_{ar}^{(11)}(k;\eta,\eta')R_{ps}^{(11)}(k_1;\eta,\eta')
\nonumber\\
&&\quad+ 
\quad\Bigl(\quad a\longleftrightarrow b, \quad \quad 
\bfk \longleftrightarrow -\bfk \quad\Bigr).
\label{eq:evolve_P13}
\end{eqnarray}
Summing up equations (\ref{eq:evolve_P22}) and (\ref{eq:evolve_P13}), 
and using the fact that 
$\widehat{\mathbf{\Sigma}}_{abcd}(\eta)\,P_{cd}^{(11)}(k;\eta)=0$, we 
finally arrive at equation (\ref{eq:recovery_1loopPT}).

\section{Integral solutions }
\label{appendix:integral_sol}

In this appendix, we show that the integral expressions given 
in Section \ref{subsec:integral_solutions} are compatible with the 
closure equations (\ref{eq:DIA_eq1}) and (\ref{eq:DIA_eq2}).  
Here, we shall particularly focus on the integral expression 
(\ref{eq:integral_2}) and 
explicitly derive the closure equation (\ref{eq:DIA_eq2}) 
from equation (\ref{eq:integral_2}).  
As for the integral expression (\ref{eq:integral_1}), 
it is straightforward to show the compatibility between equations 
(\ref{eq:DIA_eq1}) and (\ref{eq:integral_1}), just repeating the 
same procedure as presented below.

Let us consider equation (\ref{eq:integral_2}) and separate 
the right-hand side of this equation into two terms:  
\begin{eqnarray}
R_{bc}^{\rm(I)}(k;\eta,\eta')&=&
G_{bd}(k|\eta,\eta_0)\,G_{ce}(k|\eta',\eta_0)\,P_{de}(k;\eta_0),
\nonumber\\
R_{bc}^{\rm(II)}(k;\eta,\eta')&=&
\int_{\eta_0}^{\eta}d\eta_1\int_{\eta_0}^{\eta'}d\eta_2\,\,
G_{bd}(k|\eta,\eta_1)\,G_{ce}(k|\eta',\eta_2)\,\Phi_{de}(k;\eta_2,\eta_1).
\nonumber
\end{eqnarray}
Acting the operator $\widehat{\mbox{\boldmath$\Lambda$}}_{ab}(\eta)$ on 
the above equations,  
with a help of equation (\ref{eq:DIA_eq3}), we have 
\begin{eqnarray}
&&\widehat{\mathbf{\Lambda}}_{ab}(\eta)\,\,R_{bc}^{\rm(I)}(k;\eta,\eta') = 
4 \int_{\eta_0}^{\eta}d\eta_1\int\frac{d^3\bfq}{(2\pi)^3}\,
\gamma_{apq}(\bfq,\bfk-\bfq)\,\gamma_{lrs}(-\bfq,\bfk)
\nonumber
\\
&&\quad\quad\quad\quad\quad\times \,\,G_{ql}(|\bfk-\bfq||\eta,\eta_1)\,
R_{pr}(q;\eta,\eta_1)\,G_{sb}(k|\eta_1,\eta_0)\,G_{cd}(k|\eta',\eta_0)\,
P_{bd}(k;\eta_0),
\nonumber
\\
&&\widehat{\mathbf{\Lambda}}_{ab}(\eta)\,\,R_{bc}^{\rm(II)}(k;\eta,\eta')
= \int_{\eta_0}^{\eta'} d\eta_2\,\, G_{ce}(k|\eta',\eta_2)\,
\Phi_{ae}(k;\eta_2,\eta)
\nonumber
\\
&&\quad\quad
+\,\,4\int_{\eta_0}^{\eta}d\eta_1 \int_{\eta_0}^{\eta'}d\eta_2 
\int_{\eta_1}^{\eta}d\eta_3 
\int\frac{d^3\bfq}{(2\pi)^3}\,\,
\gamma_{apq}(\bfq,\bfk-\bfq)\,\gamma_{lrs}(-\bfq,\bfk),
\nonumber
\\
&&\quad\quad\quad\quad
\times \,\,G_{ce}(k|\eta',\eta_2)\,\Phi_{de}(k;\eta_2,\eta_1)\,
G_{ql}(|\bfk-\bfq||\eta,\eta_3)\,
R_{pr}(q;\eta,\eta_3)\,G_{sd}(k|\eta_3,\eta_1).  
\nonumber
\end{eqnarray}
Summing up the above two equations, we obtain
\begin{eqnarray}
\widehat{\mathbf{\Lambda}}_{ab}(\eta)\,\,
\left\{\,R_{bc}^{\rm(I)}+R_{bc}^{\rm(II)}\right\}
&=&\int_{\eta_0}^{\eta'} d\eta_2\,\, G_{ce}(k|\eta',\eta_2)\,
\Phi_{ae}(k;\eta_2,\eta)
\nonumber\\
&&\quad\quad
+\,4\int\frac{d^3\bfq}{(2\pi)^3}\,\,
\gamma_{apq}(\bfq,\bfk-\bfq)\,\gamma_{lrs}(-\bfq,\bfk)\,
\Bigl[~\cdots~\Bigr],  
\label{appendix_inter_step}
\end{eqnarray}
where the bracket $[\,\,\cdots\,\,]$ in the second term of right-hand side 
is rewritten as 
\begin{eqnarray}
\Bigl[~\cdots~\Bigr] &=&
\int_{\eta_0}^{\eta}d\eta_1 
G_{ql}(|\bfk-\bfq||\eta,\eta_1)\,
R_{pr}(q;\eta,\eta_1)\,G_{sb}(k|\eta_1,\eta_0)\,G_{cd}(k|\eta',\eta_0)\,
P_{bd}(k;\eta_0)
\nonumber\\
&&+\int_{\eta_0}^{\eta}d\eta_1 \int_{\eta_0}^{\eta'}d\eta_2 
\int_{\eta_1}^{\eta}d\eta_3 \,
G_{ce}(k|\eta',\eta_2)\,\Phi_{de}(k;\eta_2,\eta_1)\,
\nonumber\\
&&\quad\quad\quad\quad\quad\quad\quad\quad\quad\times\,\,
G_{ql}(|\bfk-\bfq||\eta,\eta_3)\,
R_{pr}(q;\eta,\eta_3)\,G_{sd}(k|\eta_3,\eta_1)  
\nonumber\\
&=& \int_{\eta_0}^{\eta}d\eta_1\,\, G_{ql}(|\bfk-\bfq||\eta,\eta_1)\,
R_{pr}(q;\eta,\eta_1)
\nonumber\\
&&\times \Bigl[\,\, 
G_{sb}(k|\eta_1,\eta_0)\,G_{cd}(k|\eta',\eta_0)\,
P_{bd}(k;\eta_0)
\Bigr.
\nonumber\\
&&\quad\Bigl.+ \int_{\eta_0}^{\eta}d\eta_2 \int_{\eta_0}^{\eta'}d\eta_3\,\,
\Theta(\eta_1-\eta_2)\,G_{sd}(k|\eta_1,\eta_2)\,G_{ce}(k|\eta',\eta_3)\,
\Phi_{de}(k;\eta_3,\eta_2)\,\,
  \Bigr]
\nonumber\\
&=& \int_{\eta_0}^{\eta}d\eta_1\,\, G_{ql}(|\bfk-\bfq||\eta,\eta_1)\,
R_{pr}(q;\eta,\eta_1)
\nonumber\\
&&\quad\quad\quad\quad\quad\quad\quad
\times \Bigl[\,\, R_{sc}(k;\eta_1,\eta')\,\Theta(\eta_1-\eta')+
R_{cs}(k;\eta',\eta_1)\,\Theta(\eta'-\eta_1)\,\,\Bigr].
\nonumber
\end{eqnarray}
Note that, in the second equality, integration variables
$(\eta_1,\eta_2,\eta_3)$ has been periodically replaced with 
$(\eta_2,\eta_3,\eta_1)$ and the domain of integral for $\eta_1$ were 
expanded by introducing the Heaviside step function.
On the other hand, with a help of the expression $\Phi_{ab}$
(see Eq.[\ref{eq:mode_coupling}]), 
the first term in equation (\ref{appendix_inter_step}) becomes
\begin{eqnarray}
\int_{\eta_0}^{\eta'} d\eta_2\,\, G_{ce}(k|\eta',\eta_2)\,
\Phi_{ae}(k;\eta_2,\eta)
&=&\,2\int\frac{d^3\bfq}{(2\pi)^3}\,
\gamma_{apq}(\bfq,\bfk-\bfq)\,\gamma_{lrs}(\bfq,\bfk-\bfq)\,
\nonumber\\
& \times&
\int_{\eta_0}^{\eta'}d\eta_1\,
G_{cl}(k|\eta',\eta_1)\,
R_{pr}(q;\eta,\eta_1)\,R_{qs}(|\bfk-\bfq|;\eta,\eta_1).
\nonumber
\end{eqnarray}

Now, collecting these results, equation (\ref{appendix_inter_step}) becomes 
\begin{eqnarray}
&&\widehat{\mathbf{\Lambda}}_{ab}(\eta)\,\,
\left\{\,R_{bc}^{\rm(I)}+R_{bc}^{\rm(II)}\right\}
\nonumber\\
&&\quad\quad
=\int\frac{d^3\bfq}{(2\pi)^3}\,\,\gamma_{apq}(\bfq,\bfk-\bfq)\,\,
\nonumber\\
&&\quad\quad\quad\quad
\times\,\,\Bigl[\,\,
4\int_{\eta_0}^{\eta}d\eta_1\,\,\gamma_{lrs}(-\bfq,\bfk)\,
G_{ql}(|\bfk-\bfq||\eta,\eta_1)\,R_{pr}(q;\eta,\eta_1)
\Bigr.
\nonumber\\
&&\quad\quad\quad\quad\quad\quad\quad\quad\quad\times\,\,\Bigl\{
R_{sc}(k;\eta_1,\eta')\,\Theta(\eta_1-\eta')+
R_{cs}(k;\eta',\eta_1)\,\Theta(\eta'-\eta_1)\Bigr\}
\nonumber\\
&& \Bigl.\quad\quad\quad+\,\,
2\int_{\eta_0}^{\eta'}d\eta_1\,\,\gamma_{lrs}(\bfq,\bfk-\bfq)\,
G_{cl}(k|\eta',\eta_1)\,R_{pr}(q;\eta,\eta_1)
R_{qs}(|\bfk-\bfq|;\eta,\eta_1)
\,\,\Bigr]
\nonumber\\
&&\quad\quad
=\,\int\frac{d^3\bfq}{(2\pi)^3}\,\,\gamma_{apq}(\bfq,\bfk-\bfq)\,\,
K_{cpq}(-\bfk,\bfq,\bfk-\bfq;\eta,\eta'),
\nonumber
\end{eqnarray}
with the use of the definition (\ref{eq:kernel_K2}). This is exactly 
the same form as in the closure equation (\ref{eq:DIA_eq2}) and 
in this sense, the 
the integral expression (\ref{eq:integral_2}) is compatible with 
closure equation.

\section{Non-linear propagator}
\label{appendix:func_propagator}

Here, we summarize the explicit expressions for time-dependent functions 
$\alpha$, $\beta_{g,d}$, $\gamma_{g,d}$ and $\delta$ which appear 
in the one-loop propagator $G_{ab}^{(1)}$ \citep[see also][]{CS2006b}: 
\begin{eqnarray}
\alpha(\eta)&=&e^{2\eta}-\frac{7}{5}\,e^{\eta}+\frac{2}{5}\,e^{-3\eta/2},
\nonumber\\
e^{\eta}\beta_g(\eta)&=&e^{-3\eta/2}\beta_d(\eta)
=\frac{3}{5}\,e^{2\eta}-e^{\eta}+\frac{2}{5}\,e^{-\eta/2},
\nonumber\\
e^{\eta}\gamma_g(\eta)&=&e^{-3\eta/2}\gamma_d(\eta)=
\frac{2}{5}\,e^{2\eta}-e^{\eta/2}+\frac{3}{5}\,e^{-\eta/2},
\nonumber\\
\delta(\eta)&=&\frac{2}{5}\,e^{7\eta/2}-\frac{7}{5}\,e^{\eta}+1.
\nonumber
\end{eqnarray}
Further, we list the explicit expressions for the scale-dependent 
functions $f$, $g$, $h$ and $i$ \citep[][]{CS2006b}: 
\begin{eqnarray}
f(k)&=&\frac{1}{504} \int\frac{d^3\bfq}{(2\pi)^3\,k^3q^5}\,
\Bigl\{ 6 k^7q- 79 k^5q^3 + 50 k^3q^5 - 21 kq^7 \Bigr.
\nonumber\\
&&\quad\quad\quad\quad\quad\quad\quad\quad\quad\quad\quad\quad
\Bigl.+
\frac{3}{4}\left(k^2-q^2\right)^3\left(2k^2+7q^2\right)
 \ln \left|\frac{k-q}{k+q}\right|^2
\Bigr\} P_0(q),
\nonumber\\
g(k)&=&\frac{1}{168} \int\frac{d^3\bfq}{(2\pi)^3\,k^3q^5}\,
\Bigl\{ 6k^7q-41k^5q^3+2k^3q^5 -3kq^7 \Bigr.
\nonumber\\
&&\quad\quad\quad\quad\quad\quad\quad\quad\quad\quad\quad\quad
\Bigl.+ 
\frac{3}{4}(k^2-q^2)^3\left(2k^2+q^2\right)
\ln \left|\frac{k-q}{k+q}\right|^2
\Bigr\} P_0(q),
\nonumber\\
h(k)&=&\frac{1}{24} \int\frac{d^3\bfq}{(2\pi)^3\,k^3q^5}\,
\Bigl\{ 6 k^7q + k^5q^3 + 9kq^7 
\Bigr.
\nonumber\\
&&\quad\quad\quad\quad\quad\quad\quad\quad\quad
\Bigl.+ 
\frac{3}{4}(k^2-q^2)^2\left(2 k^4+5k^2q^2+3q^4\right)
\ln \left|\frac{k-q}{k+q}\right|^2
\Bigr\} P_0(q), 
\nonumber\\
i(k)&=&-\frac{1}{72} \int\frac{d^3\bfq}{(2\pi)^3\,k^3q^5}\,
\Bigl\{ 6k^7q+29k^5q^3-18k^3q^5+27kq^7 \Bigr.
\nonumber\\
&&\quad\quad\quad\quad\quad\quad\quad\quad\quad
\Bigl.+ 
\frac{3}{4} (k^2-q^2)^2\left(2k^4+9k^2q^2+9q^4\right) 
\ln \left|\frac{k-q}{k+q}\right|^2
\Bigr\} P_0(q).
\nonumber
\end{eqnarray}
Note again that the quantity $P_0(q)$ is the linearly extrapolated 
power spectrum given at the present time. 

\clearpage

\end{document}